\documentclass[article]{jss}\usepackage[]{graphicx}\usepackage[]{color}
\makeatletter
\def\maxwidth{ %
  \ifdim\Gin@nat@width>\linewidth
    \linewidth
  \else
    \Gin@nat@width
  \fi
}
\makeatother

\definecolor{fgcolor}{rgb}{0.345, 0.345, 0.345}
\newcommand{\hlnum}[1]{\textcolor[rgb]{0.686,0.059,0.569}{#1}}%
\newcommand{\hlstr}[1]{\textcolor[rgb]{0.192,0.494,0.8}{#1}}%
\newcommand{\hlcom}[1]{\textcolor[rgb]{0.678,0.584,0.686}{\textit{#1}}}%
\newcommand{\hlopt}[1]{\textcolor[rgb]{0,0,0}{#1}}%
\newcommand{\hlstd}[1]{\textcolor[rgb]{0.345,0.345,0.345}{#1}}%
\newcommand{\hlkwa}[1]{\textcolor[rgb]{0.161,0.373,0.58}{\textbf{#1}}}%
\newcommand{\hlkwb}[1]{\textcolor[rgb]{0.69,0.353,0.396}{#1}}%
\newcommand{\hlkwc}[1]{\textcolor[rgb]{0.333,0.667,0.333}{#1}}%
\newcommand{\hlkwd}[1]{\textcolor[rgb]{0.737,0.353,0.396}{\textbf{#1}}}%

\usepackage{framed}
\makeatletter
\newenvironment{kframe}{%
 \def\at@end@of@kframe{}%
 \ifinner\ifhmode%
  \def\at@end@of@kframe{\end{minipage}}%
  \begin{minipage}{\columnwidth}%
 \fi\fi%
 \def\FrameCommand##1{\hskip\@totalleftmargin \hskip-\fboxsep
 \colorbox{shadecolor}{##1}\hskip-\fboxsep
     \hskip-\linewidth \hskip-\@totalleftmargin \hskip\columnwidth}%
 \MakeFramed {\advance\hsize-\width
   \@totalleftmargin\z@ \linewidth\hsize
   \@setminipage}}%
 {\par\unskip\endMakeFramed%
 \at@end@of@kframe}
\makeatother

\definecolor{shadecolor}{rgb}{.97, .97, .97}
\definecolor{messagecolor}{rgb}{0, 0, 0}
\definecolor{warningcolor}{rgb}{1, 0, 1}
\definecolor{errorcolor}{rgb}{1, 0, 0}
\newenvironment{knitrout}{}{} 

\usepackage{alltt}
\usepackage{graphicx}
\usepackage{xcolor}
\usepackage{tikz}
\usetikzlibrary{calc,shapes,backgrounds,arrows,automata,shadows,positioning}
\usepackage{stmaryrd}
\usepackage{amsfonts, amssymb, dsfont,amsmath}
\usepackage{multirow}
\usepackage{multicol}
\usepackage{booktabs}
\usepackage{enumerate}
\usepackage{todonotes}

\usepackage[normalem]{ulem}

\usepackage{tikz}
\usetikzlibrary{shapes}
\usepackage{pgfplots}
\usepackage{lscape}

\def\addlegendimage{\csname pgfplots@addlegendimage\endcsname}

\pgfkeys{/pgfplots/number in legend/.style={
        /pgfplots/legend image code/.code={
            \node at (0.295,-0.0225){#1};
        },
    },
}
\newcommand{\indep}{\rotatebox[origin=c]{90}{$\models$}}

\newcommand{\MA}{Y}

\newcommand{\bX}{\mathbf{X}}
\newcommand{\MAO}{\bY^\text{\rm o}}
\newcommand{\MAM}{\bY^\text{\rm m}}

\newcommand{\btau}{\boldsymbol{\tau}}
\newcommand{\bnu}{\boldsymbol{\nu}}
\newcommand{\bZ}{\mathbf{Z}}

\newcommand{\bY}{\mathbf{Y}}

\newcommand{\bR}{\mathbf{R}}

\newcommand{\block}{\mathcal{Q}}
\newcommand{\N}{\mathcal{N}}

\newcommand{\btheta}{\boldsymbol{\theta}}
\newcommand{\bpi}{\boldsymbol{\pi}}
\newcommand{\bal}{\boldsymbol{\alpha}}
\newcommand{\bpsi}{\boldsymbol{\psi}}

\newcommand{\bbeta}{\boldsymbol{\beta}}

\newcommand{\dyad}{\mathcal{D}}

\newcommand{\dyadM}{\dyad^\text{\rm m}}

\newcommand{\node}{\mathcal{N}}

\newcommand{\Pbb}{\mathbb{P}}
\newcommand{\Ebb}{\mathbb{E}}

\newcommand{\Rbb}{\mathbb{R}}

\newcommand{\argmax}{\mathop{\mathrm{arg\ max}}}

\usepackage{amsthm}
\theoremstyle{plain}

\newtheorem{proposition}{Proposition}

\newtheorem*{properties*}{Properties}
\newtheorem*{lemma*}{Lemma}

\graphicspath{{./figures/}}

\makeatletter
\def\maxwidth{ %
\ifdim\Gin@nat@width>\linewidth
\linewidth
\else
\Gin@nat@width
\fi
}
\makeatother

\definecolor{shadecolor}{rgb}{.97, .97, .97}
\definecolor{messagecolor}{rgb}{0, 0, 0}
\definecolor{warningcolor}{rgb}{1, 0, 1}
\definecolor{errorcolor}{rgb}{1, 0, 0}

\definecolor{rouge}{RGB}{204,45,24}

\usepackage{alltt}

\usepackage{thumbpdf,lmodern}

\usepackage{framed}

\newcommand{\new}[1]{#1}

\author{Pierre Barbillon\\UMR MIA-Paris, Universit\'e Paris-Saclay, AgroParisTech, INRAE
\AND Julien Chiquet\\UMR MIA-Paris, Universit\'e Paris-Saclay, AgroParisTech, INRAE
\AND Tabouy Timoth\'ee\\UMR MIA-Paris, Universit\'e Paris-Saclay, AgroParisTech, INRAE}
\Plainauthor{Pierre Barbillon, Julien Chiquet, Timoth\'ee Tabouy}

\title{\pkg{missSBM}: An \proglang{R} Package for Handling Missing Values in the Stochastic Block Model}
\Plaintitle{missSBM: An R Package for Handling Missing Values in the Stochastic Block Model}
\Shorttitle{\pkg{missSBM}: Handling Missing Values in the SBM}

\Abstract{The Stochastic Block Model (SBM) is a popular probabilistic model for random graphs. It is commonly used for clustering network data by aggregating nodes that share similar connectivity patterns into blocks. When fitting an SBM to a network which is partially observed, it is important to take into account the underlying process that \new{generates} the missing values, otherwise the inference may be biased. This paper introduces \pkg{missSBM}, an \proglang{R}-package fitting the SBM when the network is partially observed, i.e., the adjacency matrix contains not only 1's or 0's encoding presence or absence of edges but also \texttt{NA}'s encoding missing information between pairs of nodes. This package implements a \new{set} of algorithms for fitting the binary SBM, possibly in the presence of external covariates, by performing variational inference adpated to several observation processes. Our implementation automatically explores different block numbers to select the most relevant model according to the Integrated Classification Likelihood (ICL) criterion. The ICL criterion can also help  determine which observation process better corresponds to a given dataset. Finally, \pkg{missSBM} can be used to perform imputation of missing entries in the adjacency matrix. We illustrate the package on a network data set consisting of interactions between political blogs sampled during the French presidential election in 2007.
}

\Plainkeywords{Network, Missing data, Stochastic Block Model}

\Address{
  Pierre Barbillon, Julien Chiquet \& Timoth\'ee Tabouy\\
  MIA Paris, Université Paris-Saclay, AgroParisTech, INRAE \\
  E-mails: \email{pierre.barbillon@agroparistech.fr}, \email{julien.chiquet@inrae.fr}, \\\email{timothee.tabouy@gmail.com}
}

\IfFileExists{upquote.sty}{\usepackage{upquote}}{}
\IfFileExists{upquote.sty}{\usepackage{upquote}}{}
\begin{document}


\section[Introduction]{Introduction} \label{sec:intro}

In many fields of science, networks are a natural way to represent interaction data. To cite a few examples, a network may represent social interactions such as friendship or collaboration between people in a social network, regulation between genes and their products in a gene regulatory network, or predation between animals in a food web. In this paper, we only consider networks which can be represented by graphs composed of binary edges connecting pairs of nodes (also referred to as \emph{dyads} in the following).

To this day, there exist many pieces of software performing network-related analyses. Unsurprisingly, the \proglang{R} community is extremely active in this area. Indeed,  the \proglang{R} programming language is especially well-designed for performing data manipulation and visualization, and is thus appropriate for handling network data. Among the many available \proglang{R} packages related to networks, we suggest a classification into three groups\footnote{In addition to this brief typology, the interested reader may consult the CRAN task view on the related topic of graphical modeling \citep{taskCRAN_GM}.}:
\begin{enumerate}[i)]
\item Packages for representation, manipulation or visualization tasks, and packages computing descriptive statistics.
We mention non-exhaustively the following top representatives: \pkg{igraph} \citep{igraph}, \pkg{network} and \pkg{sna} \citep{network,sna}.
\item \new{Packages learning the structure of a network from an external source of data, such as \pkg{huge} \citep{huge}, \pkg{glasso} \citep{glasso}, \pkg{bnlearn} \citep{scutari2009learning} or \pkg{bnstruct} \citep{bnstruct}. These packages generally rely on a specific graphical modeling of the data (e.g., Gaussian graphical models \citep{Lauritzen1996} in \pkg{huge} and \pkg{glasso}, or Bayesian networks \citep{pearl2011bayesian} in
\pkg{bnlearn} and \pkg{bnstruct})}.
\item \new{Packages fitting (probabilistic) models on network data. The \pkg{ergm} package \citep{ergm} fits the family of exponential random graph models (ERGM) introduced in \cite{ergm_model}: it is part of the collection of tools around ERGM regrouped in the \pkg{statnet} metapackage \citep{statnet};
\pkg{latentnet}  \citep{latentnet} implements the latent space approach of \cite{hoff2002latent}; \pkg{mixer} \citep{mixer} and \pkg{blockmodels} \citep{Leger2016} fit the Stochastic Block Model (SBM) when the distribution of the edges belongs to the exponential family \citep{snijders1997estimation,Nowicki2001}. Other \proglang{R} packages related to the SBM and its extensions include \pkg{sbm} \citep{sbm}, \pkg{sbmr} \citep{sbmr}, \pkg{dynSBM} \citep{dynsbm},
 \pkg{blockmodeling} \citep{blockmodeling}, \pkg{dBlockmodeling} \citep{dBlockmodeling},
 \pkg{expSBM} \citep{expSBM}, \pkg{MLVSBM} \citep{MLVSBM}, \pkg{greed} \citep{greed},
  \pkg{sbmSDP} \citep{sbmSDP}, \pkg{hergm} \citep{schweinberger2018hergm}, \pkg{lda} \citep{lda},
 \pkg{graphon} \citep{graphon}, \pkg{GREMLINS} \citep{GREMLINS} and \pkg{noisySBM} \citep{noisySBM}. Some of these packages, as well as some implementations in other programming languages, are presented in the following.}

\end{enumerate}

The \pkg{missSBM} package which we introduce here belongs to the third category, that is, software that fits a specific probabilistic model on network data. More specifically, \pkg{missSBM} is dedicated to the estimation of the Stochastic Block Model (SBM), a mixture of Erd\H{o}s-R\'enyi random graphs  \citep{Erdos1959} offering a high degree of heterogeneity in connectivity profiles \new{\citep[see][for a recent review]{abbe2017community}}. The SBM generally fits well real-world network data while keeping the advantage of being a probabilistic generative model (contrary to  mechanistic approaches such as the Barab\'asi-Albert model \citep{albert2002statistical}, defined by a preferential attachment algorithm). The main outcome of an SBM fit is a clustering of the nodes -- or "blocks" -- so that the nodes share the same properties within the same block. To our knowledge, the reference package for fitting the SBM with the \proglang{R} programming system is \pkg{blockmodels}. It includes efficient implementations of variational algorithms to fit different flavors of the SBM, adapted to binary network data and valued networks, with optional covariates on the edges. Two other important extensions of the SBM are available as \proglang{R} packages: the degree-corrected Stochastic Block Model in \pkg{randnet} \citep{randnet} and a dynamic version of the Stochastic Block Model in \pkg{dynsbm} \citep{dynsbm}.
Beyond the \proglang{R} framework, there also exist \proglang{Python} packages and \proglang{C++} libraries providing efficient codes for some particular SBM: the \proglang{Python} packages \pkg{CommunityDetection} \citep{communityDetection} and \pkg{BipartiteSBM} \citep{bipartiteSBM} are dedicated to the estimation of special network structures using various heuristics and network models, among which the SBM. Beyond variational approaches, MCMC methods exist for inferring the SBM, solving the exact problem but being generally more computationally demanding: {the \proglang{Python} library \pkg{graph-tool} \citep{graph-tool} includes an MCMC sampler to fit the binary SBM and its degree-corrected variant;} \proglang{C++} libraries \pkg{sbm\_canonical\_mcmc} \citep{sbmCanonicalMCMC} and \pkg{bipartiteSBM-MCMC} \citep{bipartiteSbmMCMC} respectively implement a MCMC sampler for the SBM and the bipartite SBM. Finally \pkg{MODE-NET} \citep{modeNet} implements the belief propagation algorithm for inferring the degree-corrected SBM.

Despite their high quality, an important limitation of the aforementioned software is to require a network that is fully observed, that is, no missing value is supported. The main feature of \pkg{missSBM} is to deal with  cases where the network data is only partially observed. More precisely, we consider situations where the adjacency matrix of the network data contains not only 1's or 0's for presence or absence of an edge, but also \texttt{NA}'s encoding missing information for some dyads. \new{Note that this situation is different from the
case considered in \pkg{noisySBM}: there, a similarity matrix is fully observed between all pairs of nodes,
and the goal is to separate the 'true' interactions from noise by means of a dedicated SBM.}

\new{When inferring the SBM from network data with missing values,} it is important to take into account the underlying process that \new{generates} these missing values in the estimation of the model parameters, otherwise it  may be biased. More specifically, one has to identify whether the  values are Missing at Random or not \citep[MAR and MNAR, see][]{little2019statistical}. This issue has been studied in the context of network data by \cite{handcock2010} for the ERGM and in our methodological paper \citep{Tabouy2019} for the SBM. \pkg{missSBM} is an implementation of the methodology developed therein. It also considers new sampling designs and the inclusion of covariates simultaneously in the SBM and in the observation process, which was not studied by \citet{Tabouy2019}. Specifically, \pkg{missSBM} implements variational algorithms in the vein of \cite{daudin2008mixture} and \cite{Leger2016} for estimating the SBM, with or without covariates, under various missing data mechanisms. This includes cases of incomplete data where the inference can be made only on the observed part of the data (MAR), or cases where it is necessary to take the sampling design into account in the inference (MNAR). 

Some frameworks deal with missing data but rather from the cross-validation perspective than the sampling perspective. Cross-validation is used to perform model selection for networks such as the choice of the number of blocks or communities \citep{li2020network,chen2018network} or the choice of the latent structure \citep{hoff2008modeling}. Hence, these frameworks are quite different from ours since cross-validation is done under a MAR sampling while our main goal is to be able to infer an SBM under several MNAR sampling mechanisms.

The paper is organized as follows: Section \ref{sec:models} introduces the statistical framework of the binary SBM, with or without covariates, and summarizes the key points of its inference under missing data conditions. 
Section \ref{sec:guidelines} provides basic user guidelines for the main functions and classes of objects. We finally detail in Section \ref{sec:example} a case study which analyzes a network data set describing the French blogosphere during the period preceding the 2007 French presidential election, illustrating the most striking features of the package.

\section{Statistical Framework} \label{sec:models}

\subsection{Binary Stochastic Block Model (SBM)} \label{subsec:sbm}

In an SBM, nodes from a set $\mathcal{N} \triangleq \{ 1,\ldots,n \}$ are distributed among a set $\mathcal{Q} \triangleq \{ 1,\ldots,Q \}$  of hidden blocks which model the latent structure of the graph. The group membership is described by {independent} categorical variables $(\bZ_i, i \in \mathcal{N})$ with multinomial distribution $\mathcal{M}(1,\boldsymbol{\alpha} = ( \alpha_1,...,\alpha_Q ))$.  The probability of having an edge between any pair of nodes (or \textit{dyad}) only depends on the blocks the two nodes belong to. Hence, the presence of an edge between $i$ and $j$, indicated by the binary variable $Y_{ij}$, is independent of the other edges conditionally on the latent blocks:
\begin{equation}
\label{eq:SBM}
\MA_{ij} | \bZ_i, \bZ_j \sim^{ind} \mathcal{B}(\pi_{\bZ_i \bZ_j}), \ \text{for all}\  (i,j) \in \mathcal{D},
\end{equation}
where $\mathcal{B}$ stands for the Bernoulli distribution {and $\mathcal{D}$ the set of dyads. This set may be either equal  to $\{(i,j) \in\mathcal{N}^2;\ i\not=j\}$ if the network is directed or to $\{(i,j) \in\mathcal{N}^2;\ i<j \}$, otherwise\footnote{Although self-edges ($\MA_{ii}$) could be defined in the SBM, they are not considered in \pkg{missSBM} since they are scarce in real data.}.}
In the following, we denote by ${\boldsymbol{\pi}} = \left(\pi_{q\ell}\right)_{(q,\ell) \in \mathcal{Q}^2}\in[0,1]^{Q^2}$ the connectivity matrix, $\boldsymbol{\alpha}\in \mathbb{D}^{Q}=\{(\alpha_1,\ldots,\alpha_Q)\in[0,1]^{Q};\ \alpha_1+\ldots\alpha_Q=1 \}$ the {block proportions}, ${\bZ} = (\bZ_1,...,\bZ_n)^T$ the $n \times Q$ membership matrix and $\mathbf{Y}=(Y_{ij})_{(i,j)\in\mathcal{D}}$ the $n\times n$ adjacency matrix. This matrix is binary, with a diagonal filled with \texttt{NA}'s and is symmetric if and only if the network is undirected. The vector encompassing all the unknown model parameters is $\boldsymbol{\theta} = (\boldsymbol{\alpha}, \boldsymbol{\pi})$. A schematic representation of the binary SBM in the undirected case is given in Figure \ref{fig:tikzSBM}, where we highlight the latent clustering.

\begin{figure}[htbp!]
\centering
\begin{multicols}{2}
\begin{tikzpicture}

\tikzstyle{every edge}=[-,>=stealth',shorten >=1pt,auto,thin,draw]
\tikzstyle{every state}=[draw=none,text=white,scale=0.65, font=\scriptsize, transform shape]
\tikzstyle{every node}=[fill=yellow!40!orange]
\node[state] (A1) at (0,0.5) {A1};
\node[state] (A2) at (1,0.5) {A2};
\node[state] (A3) at (.5,1.5) {A3};

\path (A2) edge [bend left] node[fill=white,below=.1cm]
{$\pi_{\textcolor{yellow!40!orange}{\bullet}\textcolor{yellow!40!orange}{\bullet}}$}
(A1)
(A1) edge [bend left] (A3)
(A3) edge [bend left] (A2);

\tikzstyle{every node}=[fill=blue!80!black]
\foreach \angle/\text in {234/B1, 162/B2, 90/B3, 18/B4, -54/B5} {
\node[fill=blue,state,xshift=5cm,yshift=3.5cm]     (\text)    at
(\angle:1cm) {\text};
}
\path (B2) edge (B5)
(B1) edge (B4);
\foreach \from/\to in {1/2,2/3,4/5,5/1}{
\path (B\from) edge [bend left] (B\to);
}

\path    (B3)    edge     [bend    left]    node[fill=white]
{$\pi_{\textcolor{blue!80!black}{\bullet}\textcolor{blue!80!black}{\bullet}}$}  (B4) ;

\tikzstyle{every node}=[fill=green!50!black]
\node[state] (C1) at (3,-.5) {C1};
\node[state] (C2) at (4,0) {C2};

\path (C1) edge [bend right] node[fill=white,below=.25cm]
{$\pi_{\textcolor{green!50!black}{\bullet}\textcolor{green!50!black}{\bullet}}$}
(C2);

\path (A3) edge [bend right]  (B2)
(A3)    edge    [bend    left]    node[fill=white]
{$\pi_{\textcolor{yellow!40!orange}{\bullet}\textcolor{blue!80!black}{\bullet}}$}
(B3)
(C2) edge [bend right] node[fill=white,right]
{$\pi_{\textcolor{blue!80!black}{\bullet}\textcolor{green!50!black}{\bullet}}$}
(B4)
(A2) edge [bend right] node[fill=white]
{$\pi_{\textcolor{yellow!40!orange}{\bullet}\textcolor{green!50!black}{\bullet}}$}
(C1);
\end{tikzpicture}
\columnbreak
\vspace*{\stretch{1}}
\begin{itemize}
\item $\mathcal{Q}=\{\textcolor{yellow!40!orange}{\bullet},\textcolor{blue!80!black}{\bullet},\textcolor{green!50!black}{\bullet}\}$ blocks
\item $\alpha_\bullet  =  \mathbb{P}(i  \in  \bullet)$, $\bullet\in\mathcal{Q},i=1,\dots,n$
\item $\pi_{\textcolor{yellow!40!orange}{\bullet}\textcolor{blue!80!black}{\bullet}}     =      \mathbb{P}(Y_{ij}=1| i\in\textcolor{yellow!40!orange}{\bullet},j\in\textcolor{blue!80!black}{\bullet})$
\end{itemize}
\vspace*{\stretch{1}}
\end{multicols}
\caption{Schematic representation of an undirected network following the stochastic block model with $3$ blocks. Colors are blocks in which nodes are dispatched with probabilities $\boldsymbol\alpha$ and the distribution of the dyads depends on colors of nodes with probabilities $\boldsymbol\pi$.}
\label{fig:tikzSBM}
\end{figure}

\subsection{Accounting for External Covariates} \label{subsec:sbmWithCov}

On top of information about connections between nodes, it is common for network data to be accompanied with additional information on nodes or dyads, that we call \textit{covariates}: for instance in social networks, nodes may belong to different categories (gender, occupation, nationality). Covariates on dyads usually represent similarity or dissimilarity between nodes: for example in a context of spatial data where nodes correspond to entities with explicit geographic locations, dyad covariates may be the distances between the nodes.

Depending on the analysis, we may want to detect a connectivity pattern beyond the covariate effect. To do so, we implemented in \pkg{missSBM} a variant of Model \eqref{eq:SBM} for including covariates. Let  $\bX_{ij}\in\mathbb{R}^m$ denote the vector of $m$ covariates for dyad $(i,j)$. If the covariates correspond to the nodes, i.e., $\bX_i \in \Rbb^N$ is associated with node $i$ for all $i\in \mathcal{N}$, they are transferred onto the dyad level through a symmetric "similarity" function $\phi(\cdot,\cdot):\Rbb^N \times \Rbb^N \to \Rbb^m$:  $\bX_{ij} \triangleq \phi(\bX_i,\bX_j)$. In the following, $\bX \triangleq [\bX_{ij}]_{i,j\in \mathcal{N}}\in (\Rbb^m)^{n\times n}$ denotes the array of covariates. An SBM including the effect of these covariates is
\begin{equation}
\label{eq:SBM_covariates}
\begin{array}{rcl}
  \bZ_i \ & \sim^{\text{iid}} & \mathcal{M}(1,\boldsymbol{\alpha}), \quad \ \text{for all}\ i \in \mathcal{N}, \\
  \MA_{ij} \ | \ \bZ_i,\bZ_j, \bX_{ij}
& \sim^{\text{ind}} & \mathcal{B}(g(\gamma_{z_{i}z_{j}} + \boldsymbol{\beta}^\top \bX_{ij}), \quad \ \text{for all}\ (i,j) \in \mathcal{D},
\end{array}
\end{equation}
where $\gamma_{q\ell} \in \Rbb$, $\boldsymbol{\beta} \in \Rbb^m$, $\boldsymbol{\alpha} = (\alpha_1,...,\alpha_Q)$, $g(x)=(1+e^{-x})^{-1}$. The vector of unknown parameters is now defined by  $\boldsymbol{\theta} = (\boldsymbol{\gamma}, \boldsymbol{\beta}, \boldsymbol{\alpha})$.
{Note the connection with logistic regression: Model \eqref{eq:SBM_covariates} assumes a logistic link between the presence of an edge $Y_{ij}$ and the corresponding covariates. The intercept term $(\gamma_{q\ell})_{q\ell}$ depends on the blocks of the nodes and describes the heterogeneity of the connections that is not explained by the regression term $\beta^T\bX_{ij}$.}

\paragraph*{Connections with Similar Models.} The SBM was originally introduced by \citet{Nowicki2001}. Many extensions have been proposed since, including other distributions on dyads on top of the incorporation of covariates \citep{mariadassou2010}. Contrary to the latent space model of \citet{hoff2002latent} where the latent space is continuous, the latent variables in the SBM lie in a discrete space. When covariates are included as in Equation \eqref{eq:SBM_covariates}, their effect is removed and the blocks shall then explain the structure in the network \textit{beyond} the covariates. \new{This approach is similar to \citet{vu2013model} and opposite to the one used by \citet{tallberg2004bayesian} or \citet{binkiewicz2017covariate} where the covariates help learn the underlying clustering of the nodes since their distribution is assumed to depend on the same latent variables as the SBM.}

\subsection{Missing Data and SBM}

\new{The main feature of \pkg{missSBM} is to deal with some processes that generate missing values in order to provide more accurate estimates of the parameters underlying an incompletely observed network.} {The number of nodes $n$ is assumed to be known and the missing information only concerns the dyads.} Hence the sampled data can be encoded in an adjacency matrix $\bY$ where missing information -- dyads whose value is unobserved -- is encoded by \texttt{NA}'s.  We also define the $n \times n$ observation matrix $\bR$ such as $R_{ij} = 1$ if the dyad $\MA_{ij}$ is observed and $R_{ij} = 0$ otherwise. For convenience we define $ \bY^o = \{ \MA_{ij} : R_{ij} = 1 \}$ and $\bY^m = \{ \MA_{ij} : R_{ij} = 0 \}$ the respective sets of observed and unobserved dyads.

In our framework, an observation process -- or sampling design -- is a stochastic process that generates $\bR$. We then rely on the standard missing data theory of \citet{little2019statistical} to classify those designs either into Missing Completely At Random (MCAR), Missing At Random (MAR) or Missing Not At Random (MNAR) cases.
{This framework has to be extended to handle the latent variables $\bZ$ in the SBM
as we did in \cite{Tabouy2019}:}
\begin{equation}
\label{def:missingness}
\text{Sampling design for SBM is }
\begin{cases}
\text{MCAR} & \text{if } \bR \ \indep \ (\bY, \bZ) \ | \ {\bX},\\
\text{MAR} & \text{if } \bR \ \indep \ (\MAM,\bZ) \ | \ (\MAO,{\bX}),\\
\text{MNAR} & \text{otherwise}.\\
\end{cases}
\end{equation}
\new{The notation $\indep $ stands for independence between random variables.}
Note that MCAR missingness is a particular case of MAR missingness.
{This definition provides the general case when covariates $\bX$ are available. Otherwise, the definition remains valid just by removing $\bX$.}
Denoting by $\boldsymbol{\psi}$ the set of parameters associated with the distribution that generates the sampling matrix $\bR$, we assume that  $\boldsymbol{\psi}$ and $\boldsymbol{\theta}$ live in a product space, so that we can derive the following proposition:
\begin{proposition}\label{prop:mar} From
\eqref{def:missingness}, if the sampling is MAR or MCAR then
maximizing $p_{\boldsymbol{\theta}, \boldsymbol{\psi}}(\MAO,\bR,{\bX})$ or $p_{\boldsymbol{\theta}}(\MAO,{\bX})$ in
$\boldsymbol{\theta}$ is equivalent for $\boldsymbol{\theta}$.
\label{MAR}
\end{proposition}
This proposition was proven in \cite{Tabouy2019} in the absence of covariate. The generalization to handle covariates is straightforward.
In words, the inference can be conducted on the observed part of the network data when the sampling is M(C)AR without incurring any bias. In these cases, adaptating the existing algorithms for SBM inference is simple. MNAR sampling designs require, however, more refined inference strategies since the observation process has to be included in the inference.

\subsection{Examples of Sampling Designs for Networks} \label{samp:def}

This section reviews a set of stochastic processes defining sampling designs available in \pkg{missSBM}. These sampling designs may depend either on $i)$ the values of the dyads in the network; $ii)$ the latent clustering of the nodes; or $iii)$  some covariates, {via the vector of parameters $\bpsi$. All examples detailed in the following assume that the observations are independent conditionally on  $\bY$, $\bZ$ and $\bX$ (either observations of the dyads for dyad-centered sampling designs, or observations of the nodes for node-centered sampling designs). From a practical viewpoint, the sampling designs implemented in \pkg{missSBM} allow the user to either $i)$ generate new data by partially observing an existing network according to a predefined sampling design, with user-defined parameters $\boldsymbol\psi$, or $ii)$ to fit an SBM model under missing data condition, assuming that the missing entries arise from a given type of sampling design for which the unknown parameters $\boldsymbol\psi$ are estimated jointly with the SBM parameters $\btheta$}.

\subsubsection{Dyad-Centered Sampling Designs} \label{samp:dyad}

\begin{itemize}
\item  \textbf{Dyad Sampling} (MCAR): each   dyad    $(i,j)   \in\dyad$    has   the    same   probability  $\Pbb(R_{ij}=1) \triangleq \psi\in [0,1]$ to be observed.
\item \textbf{Double Standard Sampling} (MNAR):   let $\bpsi \triangleq (\rho_{1}, \rho_{0}) \in [0,1]^2$. Double standard sampling
consists in observing dyads with probabilities
\begin{equation*}
\Pbb(R_{ij}=1|Y_{ij}=1) = \rho_{1}, \qquad
\Pbb(R_{ij}=1|Y_{ij}=0) = \rho_{0}.
\end{equation*}
The probability for sampling a dyad thus intrinsically depends on the presence/absence of the corresponding edge.
{This double standard sampling is especially likely in real world applications, if it is easier to observe an existing connection than the absence of connection. For instance, in protein-protein networks, the sampling effort is more important to determine the absence of a link than its existence.}

\item \textbf{Block-Dyad Sampling} (MNAR): this sampling consists in observing all
dyads with probabilities $\bpsi \triangleq  (\psi_{q\ell})_{(q,\ell) \in \mathcal{Q}^2}\in [0,1]^{Q^2}$ depending on the underlying clustering of the network:
\begin{equation*}
\psi_{q\ell} = \Pbb(R_{ij}=1 \ | \ Z_{iq}=1, Z_{j\ell}=1 ).
\end{equation*}
\item \textbf{Covar-Dyad Sampling} (MAR): let us define $\bpsi \triangleq (\alpha, \kappa) \in \Rbb\times \Rbb^m$. Here the probability for observing a dyad is driven by the effect of a given covariate:
\begin{equation*}
\mathbb{P}(R_{ij} = 1 |\bX_{ij}) = g(\alpha + \kappa^t\bX_{ij}),
\end{equation*}
\new{where we recall that $g(x)=(1+e^{-x})^{-1}$.}
{Under this sampling, the external covariates may impact both  a connection and the ability to observe it. In this case, the sampling remains MAR provided that the covariates are available.}
\end{itemize}

\subsubsection{Node-Centered Sampling Designs}
\label{samp:node}

A node-centered sampling consists in observing some nodes sampled with probabilities given by the sampling design. Observing a node means observing all the dyads involving that node. For all $i\in\mathcal{N}$, we denote by $V_i$ the indicator variable for observing node $i$. Hence if $V_i=1$ we have {$R_{ij} = R_{ji} = 1$} for all $j\in \mathcal{N}$. {Node-centered sampling designs are likely in social sciences since a network is sampled through direct interviews. During an interview, individuals (nodes) indicate to whom they are connected. Some individuals may then indicate a connection with an individual not available for an interview. The resulting missing dyads concern dyads between individuals who were not interviewed. Even if the connection is oriented (directed network), we assume that an individual, when interviewed, provides its ingoing and outgoing connections.}

\begin{itemize}
\item  \textbf{Node Sampling} (MCAR): the probabilities for observing nodes are uniform: $\Pbb(V_{i}=1) = \psi\in [0,1]$ for all $i\in\mathcal{N}$.
\item {\textbf{Snowball sampling} (MAR): a first batch of nodes is sampled as in node sampling.
Then, a second batch is composed of the neighbors of
  the first batch (the set of nodes linked to at least a node of the first
  batch). Other batches can then be obtained through several sampling steps which are called waves.
  These successive waves are then MAR and not MCAR
since they are built on the basis of the previously observed part of
$\MA$.}
\item \textbf{Degree Sampling} (MNAR): for all node $i\in\mathcal{N}$, $\Pbb(V_{i}=1) = \rho_i$ where $( \rho_1, \dots, \rho_n ) \in [0,1]^n$ are such
that $\rho_{i} = g(a+b D_i)$ for all $i\in\mathcal{N}$ where
$\psi \triangleq (a,b) \in \mathbb{R}^2$ and $D_i = \sum_j Y_{ij}$. {This sampling may be the consequence of a situation where popular individuals are more likely to be interviewed.}
\item \textbf{Block-Node Sampling} (MNAR): this sampling consists in observing all
dyads corresponding to nodes selected with probabilities
$\psi \triangleq ( \psi_1, \dots, \psi_Q ) \in [0,1]^Q$ such that
$\psi_q = \Pbb(V_i=1 \ | \ Z_{iq}=1)$ for all
$(i,q) \in \node \times \block $. {This sampling may happen if some communities that shape the connections in the network are not equally reachable.}
\item \textbf{Covar-Node Sampling} (MAR): let $\psi \triangleq (\nu, \eta) \in \Rbb \times \Rbb^N$. The probability to observe a node is
\begin{equation*}
\mathbb{P}(V_{i} = 1 |\bX_i) = g(\nu + \eta^t \bX_i).
\end{equation*}
{In this sampling, some external information shapes the sampling process of the nodes. Even if the covariates also impact the probabilities of connection, as in the Covar-dyad sampling, the sampling design is MAR provided that the covariates are available.}

\end{itemize}

\subsection{Estimation Procedure: a Variational EM}

The SBM is a latent state space model which can be seen as a mixture model for random graphs. Therefore,
the EM algorithm \citep{Dempster1977} is the natural choice for the inference \new{since it generally proves very useful for inferring various types of mixture models}. It is based on the
evaluation of the expectation of the complete log-likelihood of the model, with respect to the conditional
distribution of the latent variables given the data. However, this expectation is intractable in the SBM
due to the structure of dependency between the latent variables $\bZ$ and the network $\bY$. In fact, it
would require to sum over all possible clusterings for all pairs of nodes, which is out of reach
even  for a moderate number of nodes or blocks. To address this shortcoming 
\new{when the network $\bY$ is fully observed,}
\cite{daudin2008mixture} introduced a \textit{variational} EM, based on the variational principles of
\cite{jordan1998introduction}. The idea is to maximize a lower bound of the log-likelihood based on an
approximation of the true conditional distribution of the latent variable $\bZ$. In the case of an SBM with 
missing data, the level of difficulty is higher since the set of latent variables encompasses
both $\bZ$ (the latent blocks) and $\MAM$ (the missing dyads). We propose here a variational distribution
of the conditional distribution $p_{\boldsymbol\theta,\boldsymbol\psi}(\bZ,\MAM|\MAO)$
where complete independence is forced on $\bZ$ and $\MAM$,
using a multinomial, respectively a Bernoulli  distribution for $\bZ$ and $\MAM$. \new{We denote by $m(\cdot)$ and $b(\cdot)$ the probability density functions of, respectively, the
multinomial and the Bernoulli distributions which gives the following expression of the variational distribution:}
\begin{equation*}
  \tilde p_{\boldsymbol\tau,\boldsymbol\nu} (\bZ,\MAM)= \tilde p_{\boldsymbol\tau} (\bZ) \ \tilde p_{\boldsymbol\nu} (\MAM)
  = \prod_{i\in\node} m(\bZ_{i};\tau_{i}) \prod_{(i,j) \in \dyadM} b(Y_{ij};\nu_{ij}),
\end{equation*}
\new{where $\boldsymbol\tau = \{\tau_i=(\tau_{i1},\ldots,\tau_{i\mathcal{Q}})\in [0,1]^{\mathcal{Q}}: \ \sum_{q=1}^{\mathcal{Q}}\tau_{iq}=1, i\in\node\}$  and $\boldsymbol\nu = \{\nu_{ij}\in[0,1], (i,j)\in\dyadM\}$} are the two sets of
variational parameters respectively associated with $\bZ$ and $\MAM$. Interestingly, $\boldsymbol\tau$'s are proxies for the posterior
probabilities of the group memberships for all nodes, and $\boldsymbol\nu$'s correspond to the imputed values of the missing dyads in the network data.
\new{This variational distribution was chosen in order to replace the intractable E-step of the EM algorithm with a tractable variational E-step.}
This approximation leads to the following lower bound $J$ of the log-likelihood, where $\text{KL}$ is the Kullback-Leibler divergence between the true conditional distribution
and its variational approximation:
\begin{equation*}
   \begin{aligned}
    \log p_{\boldsymbol\theta,\boldsymbol\psi}(\MAO,\bR) \geq \\
    J_{\boldsymbol\tau,\boldsymbol\nu,\boldsymbol\theta,\boldsymbol\psi}(\MAO,\bR)
    & \triangleq
    \log p_{\boldsymbol\theta,\boldsymbol\psi}(\MAO,\bR) - \text{KL}\big(\tilde p_{\boldsymbol\tau,\boldsymbol\nu}(\bZ,\MAM) \ \| \ p_{\boldsymbol\theta}(\bZ,\MAM \| \MAO)\big), \\
    & =  \Ebb_{\tilde       p_{\boldsymbol\tau,\boldsymbol\nu}}       \left[\log
      p_{\boldsymbol\theta,\boldsymbol\psi}(\MAO,\bR,\MAM,\bZ)\right]
    -\Ebb_{\tilde p_{\boldsymbol\tau,\boldsymbol\nu}}\left[ \log \tilde p_{\boldsymbol\tau,\boldsymbol\nu}(\bR,\MAM)\right].
  \end{aligned}
\end{equation*}
If we choose $\tilde p=p_{\boldsymbol\theta,\boldsymbol\psi}(\bZ,\MAM|\MAO)$, the true conditional distribution of the latent variables $\bZ,\MAM$, we retrieve the standard EM algorithm, requiring the evaluation of the intractable quantity $\Ebb_{p_{\boldsymbol\theta,\boldsymbol\psi}(\bZ,\MAM|\MAO)}       \left[\log p_{\boldsymbol\theta,\boldsymbol\psi}(\MAO,\bR,\MAM,\bZ)\right]$. Note that we alleviated the notations above by not explicitly writing the possible conditioning on covariates in the log-likelihoods.

Based on this approximation, the variational EM algorithm consists in alternating updates of the variational parameters $\{\btau, \bnu \}$ (the VE-step) with updates of the model parameters $\btheta, \bpsi$  (the M-step) maximizing $J$. Steps VE and M are iterated until convergence like in a standard EM. The algorithm converges to a local maximum of the lower bound of the log-likelihood. This variational is translated into a collection of algorithms for handling missing data with all sampling designs introduced in Section~\ref{samp:def}. When an algorithm reaches convergence, we obtain estimates of the parameters involved in the SBM ($\btheta$), in the sampling process $(\bpsi)$, and also estimates of the variational parameters which bring information on the clustering ($\btau$) and on the missing dyads ($\bnu$). The variational estimator in the SBM is proven to be asymptotically normal in \cite{bickel2013asymptotic} when the network is fully observed. The extension to the MCAR case is proven in \cite{mariadassou2019consistency}.

\paragraph*{Initialization.}

\new{It is well known that EM-like algorithms have a great sensitivity to the initialization step, which therefore requires a special attention. In \pkg{missSBM}, the initial clustering is obtained by applying the popular Absolute Eigenvalues Spectral Clustering \citep[detailed in][]{BinYu2010} to the adjacency matrix where the \texttt{NA}'s are replaced by zero. The initial clustering can also be provided by the user. As specified in the next paragraph, the exploration of the number of blocks also provides several other relevant initializations.}

\paragraph*{Selection of the Number of Blocks.}

A main difficulty met when conducting SBM inference lies in the estimation of the number of blocks, generally unknown to the user. To remedy this problem, we use the Integrated Classification Likelihood (ICL) criterion defined in \cite{Biernacki2000} and routinely used in the framework of mixture models.
\new{Note that \citet{saldana2017many}, \citet{wang2017likelihood}, \citet{hu2020using} and \citet{come2021hierarchical} propose alternative methods for selecting the number of blocks.}
{The ICL criterion was adapted to the SBM under missing data condition in \citet{Tabouy2019}, and is recalled here: for a model with $Q$ blocks, a sampling design parametrized by $K$ parameters (size of $\bpsi$) and $(\hat{\! \btheta},\hat{\bpsi})=\argmax_{(\btheta, \bpsi)}\log p_{\btheta, \psi}(\MAO,\MAM, \bR, \bZ)$, then}
{\begin{gather*}
  \mathrm{ICL}(Q)        =       -2\mathbb{E}_{\tilde        p_{\btau,\bnu}}\left[\log p_{\hat{\btheta},\hat{\bpsi}}(\MAO,\MAM, \bR, \bZ | Q, K)\right] + \mathrm{pen}_{\text{ICL}}(Q,K),\\[2ex]
  \mathrm{pen}_{\text{ICL}} = \left\{
    \begin{array}{ll}
      \left(K + \frac{Q(Q+1)}{2}\right)\log \left(\frac{n(n-1)}{2}\right) + (Q-1)\log (n) &  \text{for dyad-centered sampling,} \\
      \frac{Q(Q+1)}{2}\log\left( \frac{n(n-1)}{2} \right)+ (K + Q-1)\log (n) & \text{for node-centered sampling.}\\
    \end{array}
  \right.
\end{gather*}
}

\new{We also implemented in \pkg{missSBM} an exploration procedure designed to avoid getting stuck in local minima by producing a convex and robust ICL curve. This exploration procedure is divided into two steps, forward and backward: the forward step creates new initializations for each number $Q$ of block considered, by splitting blocks obtained from estimations with $Q-1$ blocks. On the other hand, the backward step tries new initializations for each $Q$ by merging groups of the model with $Q+1$ groups. The best model in terms of ICL is always retained. The procedure can be iterated until a satisfying shape of the ICL curve is met.}

{\paragraph*{Implementation Details.} \pkg{missSBM} adopts an oriented-object programming spirit for representing most models by means of R6-classes and the \pkg{R6} package of \cite{R6}, which paves the way for future extensions. This approach is non-visible to the user, who essentially only has to deal with classical \proglang{R} functions. The time consuming pieces of code are written in \texttt{C++} using the \pkg{armadillo} library for linear algebra \citep{sanderson2016armadillo}, in conjunction with the \pkg{Rcpp} and \pkg{RcppArmadillo} packages \citep{Rcpp,rcpparmadillo} to interface \texttt{C++} with \proglang{R}. The numerical performance of our implementation is of the same order as existing variational implementations of the binary SBM (like \pkg{blockmodels}). It allows to deal with networks with up to a couple of thousands of nodes. To give an idea to the reader, Figure~\ref{fig:timings} reports the timings for adjusting an SBM  to the network data considered in Section~\ref{sec:example} (the 2007 French political blogosphere, 194 nodes), for a varying number of blocks with a single Intel Core i9-9900 CPU at 3.10GHz, replicated 50 times per block-size. We report in the discussion some interesting tracks for improving the speed in the future and eventually tackle larger networks.}

\begin{knitrout}
\definecolor{shadecolor}{rgb}{0.969, 0.969, 0.969}\color{fgcolor}\begin{figure}[htbp!]

{\centering \includegraphics[width=.7\textwidth]{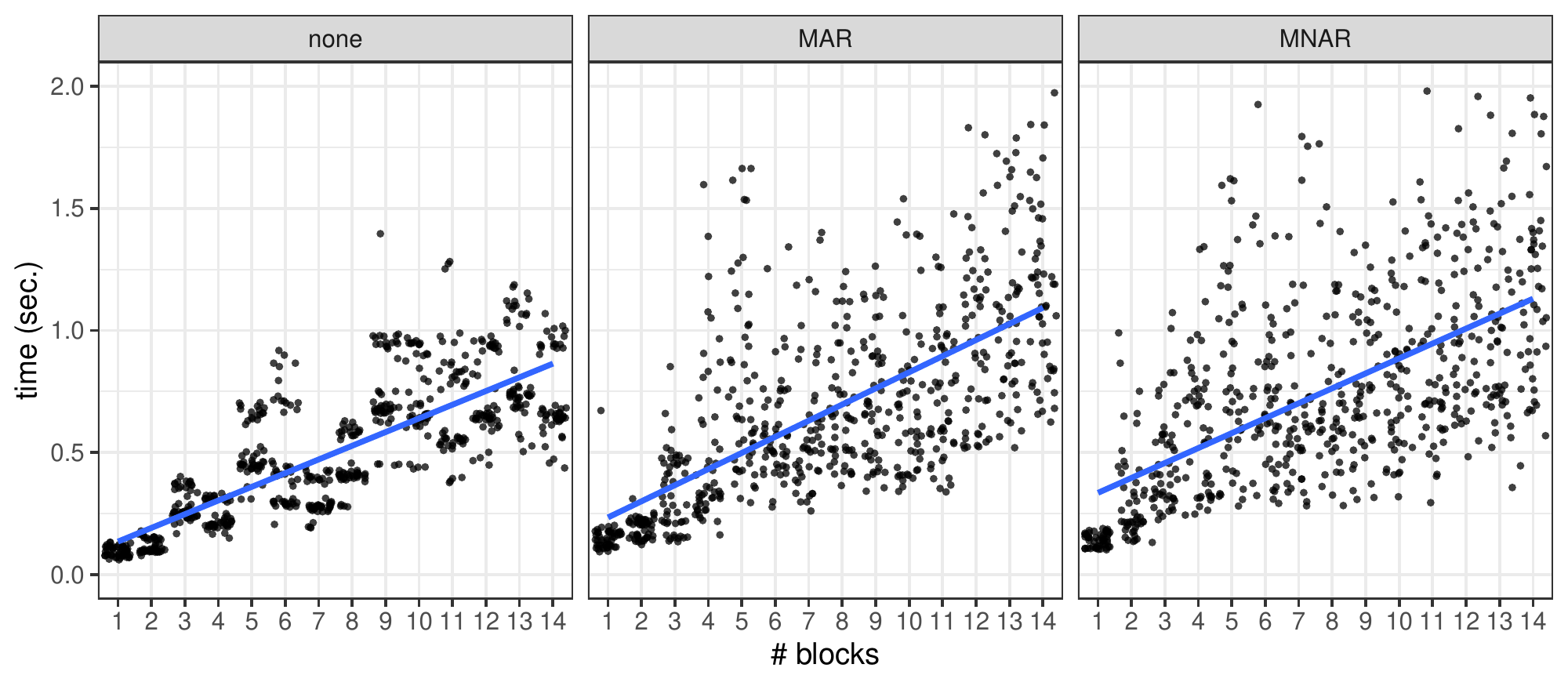} 

}

\caption{\new{\label{fig:timings}Timings for adjusting binary SBM with \pkg{missSBM} on the French political blogosphere with a single core for a varying number of blocks (50 replicated runs per \# blocks).}}\label{fig:timings}
\end{figure}

\end{knitrout}

\section{Guidelines for Users} \label{sec:guidelines}

{This section gives an overview of the basic usage of the package. In particular, it describes inputs and outputs of the main functions at play in the procedure that fits an SBM from partially observed network data. Along the section, we also describe the classes of objects included in the package to ease the manipulation of the resulting fitted models. As a complement to this section and to the usual \texttt{R} package documentation, a full documentation including a vignette is available as a \pkg{pkgdown} website at URL \url{http://grosssbm.github.io/missSBM}.}

\subsection{Parameter Estimation, Prediction and Clustering}
\label{sec:estimate}

{Estimation of an SBM from a partially observed network is done by means of the function \texttt{estimateMissSBM}, with the following usage:}

\begin{knitrout}
\definecolor{shadecolor}{rgb}{0.969, 0.969, 0.969}\color{fgcolor}\begin{kframe}
\begin{alltt}
\hlkwd{estimateMissSBM}\hlstd{(}
  \hlstd{adjacencyMatrix    ,} \hlcom{# observed network data: matrix with NA entries}
  \hlstd{vBlocks            ,} \hlcom{# vector of block numbers for model exploration}
  \hlkwc{sampling}   \hlstd{=} \hlstr{"dyad"}\hlstd{,} \hlcom{# string describing the sampling design}
  \hlkwc{covariates} \hlstd{=} \hlkwd{list}\hlstd{(),} \hlcom{# optional list of covariates (dyad or nodal)}
  \hlkwc{control}    \hlstd{=} \hlkwd{list}\hlstd{())} \hlcom{# optional list for tuning the algorithm}
\end{alltt}
\end{kframe}
\end{knitrout}

\paragraph*{Standard Usage.} This function takes as a first argument a square \code{base::matrix} \new{or a sparsely encoded \texttt{Matrix::dgCMatrix}}, possibly with \code{NA} entries corresponding to missing (unobserved) values of the network. The second argument \code{vBlocks} contains the successive explored values for the number of blocks, this number being generally unknown. The third argument -- \code{sampling} -- specifies how \code{NA} entries are taken into account in the estimation. It must be one of the following character strings, corresponding to one of the sampling designs (or observation processes) depicted in Section~\ref{samp:def}:

\begin{knitrout}
\definecolor{shadecolor}{rgb}{0.969, 0.969, 0.969}\color{fgcolor}\begin{kframe}
\begin{alltt}
\hlstd{missSBM}\hlopt{:::}\hlstd{available_samplings}
\end{alltt}
\begin{verbatim}
## [1] "dyad"            "covar-dyad"      "node"            "covar-node"     
## [5] "block-node"      "block-dyad"      "double-standard" "degree"         
## [9] "snowball"
\end{verbatim}
\end{kframe}
\end{knitrout}

{Argument \code{covariates} is an optional list with as many entries as covariates. If the covariates are \new{nodal}, each entry must be a size-$n$ vector; if the covariates \new{are defined between pairs of nodes}, each entry must be an $n\times n$ matrix.}

{\paragraph*{Advanced Tuning.} The argument \texttt{control} is a \texttt{list} to control the estimation and finely tune the variational EM algorithm, with the following entries:
\begin{enumerate}[i)]
\item \texttt{useCov}: Logical indicating whether the covariates should be incorporated within the SBM (or just in the sampling);
\item \new{\texttt{clusterInit}: Initial method for clustering. Either a character ("spectral") or a list with \code{length(vBlocks)} vectors, each with size  \code{ncol(adjacencyMatrix)}, providing a user-defined clustering. Default is "spectral", for absolute spectral clustering;}
\item \texttt{similarity}: An $\mathbb{R} \times  \mathbb{R} \mapsto \mathbb{R}$ function to compute similarities between \new{nodal} covariates (default is \texttt{missSBM:::l1\_similarity}, that is, $(x,y)\mapsto-|x-y|$);
\item \texttt{threshold}: Optimization stops when a V-EM step changes the objective function or the connection parameters by less than \texttt{threshold} (default is $1.10^{-2}$);
\item \texttt{maxIter}: Optimization stops when the number of iteration exceeds \texttt{maxIter} (default is $50$);
\item \texttt{fixPointIter}: Number of iterations in the fixed-point algorithm used to solve the variational E step (default is $3$);
\item \new{\texttt{exploration}: Character indicating what kind of exploration should be used among \texttt{"forward"}, \texttt{"backward"}, \texttt{"both"} or \texttt{"none"} (default is \texttt{"both"});}
\item \new{\texttt{iterates}: Integer for the number of iterations during the exploration process. Only relevant when \texttt{exploration} is different from \texttt{"none"} (default is 1);}
\item \new{\texttt{trace}: Logical for verbosity (default is \texttt{TRUE}).}
\end{enumerate}
}

\paragraph*{Return Value: Classes \texttt{missSBM\_collection} and \texttt{missSBM\_fit}.} An output of the function \texttt{estimateMissSBM}
is an instance of an \pkg{R6} object with class \texttt{missSBM\_collection}, which can be handled as a standard \texttt{list}. Its fields
are described in Table~\ref{tab:fieldsestimate}.
\begin{table}[htbp!]
\centering
\begin{tabular}{ll}
\hline \textbf{\small Field} & \textbf{\small Description} \\ \hline
{\small \texttt{models}} & {\footnotesize a \texttt{list} of models with class \texttt{missSBM\_fit}} \\
{\small \texttt{ICL}} & {\footnotesize the \texttt{vector} of ICL associated to the models in the collection} \\
{\small \texttt{bestModel}} & {\footnotesize best model according to the ICL (a \texttt{missSBM\_fit} object)} \\
{\small \texttt{optimizationStatus}} & {\footnotesize  a \texttt{data.frame} summarizing the optimization process for all models} \\ 
\hline \new{\textbf{\small Method}} & \new{\textbf{\small Description}} \\ \hline
\new{{\small \texttt{plot(object, type)}}} & \new{\footnotesize plot either \texttt{"icl"}, \texttt{"elbo"} or \texttt{"monitoring"}}\\ \hline
\end{tabular}
\caption{Structure of \texttt{missSBM\_collection} (output of \texttt{estimateMissSBM}).}
\label{tab:fieldsestimate}
\end{table}

Among the fields of \texttt{missSBM\_collection}, \texttt{models} is a list with as many elements as in \texttt{vBlocks}. These elements are \pkg{R6} objects with class \texttt{missSBM\_fit}, the fields of which are detailed in Table~\ref{tab:fieldsmodels}. It gives access to the results of the inference for a fixed number of blocks.
\begin{table}[htbp!]
\centering
\begin{tabular}{lll}
\hline \textbf{\small Field} & \textbf{\small Description}  & \textbf{\small \pkg{R6} object / class} \\ \hline
{\small \texttt{fittedSBM}} & {\footnotesize the adjusted Stochastic Block Model} & {\footnotesize \texttt{simpleSBM\_fit\_missSBM}} \\
{\small \texttt{fittedSampling}} & {\footnotesize the estimated sampling process} & {\footnotesize \texttt{networkSampling}} \\
{\small \texttt{imputednetwork}} & {\footnotesize the adjacency matrix with imputed values} & {\footnotesize \texttt{dgCMatrix}} \\
{\small \texttt{monitoring}}     & {\footnotesize status of the optimization process} & {\footnotesize \texttt{data.frame}} \\
{\small \texttt{vICL}} & {\footnotesize ICL criterion associated to $Q$} & {\footnotesize \texttt{double}} \\
{\small \texttt{vBound}} & {\footnotesize value of the variational bound $\mathcal{J}_{\btau, \btheta}$ at each step} & {\footnotesize \texttt{double}} \\
{\small \texttt{vExpec}} & {\footnotesize value of $\mathbb{E}_{\tilde p_{\btau}} \left[\log(p_{\btheta}(\bY, \bZ)) \right]$ at each step} & {\footnotesize \texttt{double}} \\
{\small \texttt{penalty}} & {\footnotesize penalty of the model with $Q$ blocks} & {\footnotesize \texttt{double}} \\
\hline \textbf{\small Method} & \textbf{\small Description} & \\ \hline
{\small \texttt{plot(object, type)}} & \multicolumn{2}{l}{\footnotesize plot either \texttt{"imputed"}, \texttt{"expected"}, \texttt{"meso"} or \texttt{"monitoring"}}\\
{\small \texttt{print(object)}} & \multicolumn{2}{l}{\footnotesize a print method recalling important fields and methods available} \\
{\small \texttt{coef(object, type)}} & \multicolumn{2}{l}{\footnotesize model parameters (\texttt{type}=\texttt{"mixture"}, \texttt{"connectivity"}, \texttt{"covariates"} or \texttt{"sampling"} )} \\
{\small \texttt{predict(object)}} & \multicolumn{2}{l}{\footnotesize estimated adjacency matrix (with imputation)} \\
{\small \texttt{fitted(object)}} & \multicolumn{2}{l}{\footnotesize expected value of the SBM} \\ \hline
\end{tabular}
\caption{Selection of fields in object \texttt{missSBM\_fit} with descriptions and types.}
\label{tab:fieldsmodels}
\end{table}

{We finally give additional details on fields \texttt{fittedSBM} and \texttt{fittedSampling} in Table \ref{tab:fieldsestimate2}. Note that \texttt{fittedSBM} enjoys some standard \texttt{plot}, \texttt{print}, \texttt{coef}, \texttt{predict} and \texttt{fitted} methods, most of them inherited from the class \texttt{simpleSBM\_fit} of the package \pkg{sbm}.}
\begin{table}[htbp!]
\centering
\begin{tabular}{lllc}
\hline {\small \textbf{\pkg{R6} object}} & \quad \textbf{\small Field} & \textbf{\small Description} & \textbf{\small Correspondence} \\ \hline
{\small \texttt{}} & \vline \quad {\small \texttt{probMemberships}} & {\footnotesize estimated probability of block belonging} & {\footnotesize $\{\btau_{i}\}_{i \in \N}$} \\
{\small \texttt{}} & \vline \quad {\small \texttt{connectParam}} & {\footnotesize connectivity matrix} & {\footnotesize $\hat\bpi$}  \\
\multirow{2}{*}{\texttt{fittedSBM}} & \vline \quad {\small \texttt{expectation}} & {\footnotesize imputed adjacency matrix} & {\footnotesize $\bY^o \cup {\{\nu_{ij}\}_{(i,j) \in \mathcal{D}^m}}$} \\
& \vline \quad {\small \texttt{covarParam}} & {\footnotesize regression parameter} & {\footnotesize $\hat\bbeta$}  \\
{\small \texttt{}} & \vline \quad {\small \texttt{memberships}} & {\footnotesize vector of blocks memberships} & {\footnotesize (\texttt{which.max}($\btau_i))_{i \in \N}$}  \\
{\small \texttt{}} & \vline \quad {\small \texttt{blockProp}} & {\footnotesize block proportions} & {\footnotesize $\hat\bal$} \\ \hline
{\small \texttt{fittedSampling}} & \vline \quad {\small \texttt{parameters}} & {\footnotesize sampling parameter} & {\footnotesize $\hat\bpsi$} \\ \hline
\end{tabular}
\caption{Selection of fields in \texttt{fittedSBM}, \texttt{fittedSampling} and mathematical counterparts.}
\label{tab:fieldsestimate2}
\end{table}

\paragraph*{Models Exploration.} At the end of the estimation process, it is common that the algorithm gets stuck in some local minima for some values of $Q$, the current number of blocks. \new{The consequence is "non-smooth" ICL curve which is theoretically convex and rather smooth. This may lead the user to choose a sub-optimal number of blocks. Thus, the ICL criterion is automatically "smoothed" after a first pass across all the models. The idea is to apply a split and merge strategy to the path of models stored in \texttt{missSBM\_collection} in order to find a better initialization for each value of $Q$ in the V-EM algorithm, so that it converges to the global minimum. This model exploration can be tuned with the \texttt{control} argument, see paragraph \textit{Advanced tuning}. The default goes back and forth a single time.}

\new{\paragraph*{Parallel Computing.} Some internal components of \texttt{estimateMissSBM} (initialization, estimation, exploration) use the \pkg{future} framework \citep{future} to speed-up the whole process with parallel computing. The \texttt{future::plan} must be set by the user (default is sequential with no parallelism). For instance, in order to run \pkg{missSBM} on 10 cores with forking on Unix systems, the following command should be used before calling \texttt{estimateMissSBM}().}

\begin{knitrout}
\definecolor{shadecolor}{rgb}{0.969, 0.969, 0.969}\color{fgcolor}\begin{kframe}
\begin{alltt}
\hlstd{future}\hlopt{::}\hlkwd{plan}\hlstd{(}\hlstr{"multicore"}\hlstd{,} \hlkwc{workers} \hlstd{=} \hlnum{10}\hlstd{)}
\end{alltt}
\end{kframe}
\end{knitrout}

\subsection{Partial Observation ("sampling") of Some Network Data} \label{sec:sampling}

{The function \texttt{observeNetwork} generates missing entries in a fully observed adjacency matrix according to a given network sampling design. The usage is the following:}

\begin{knitrout}
\definecolor{shadecolor}{rgb}{0.969, 0.969, 0.969}\color{fgcolor}\begin{kframe}
\begin{alltt}
\hlkwd{observeNetwork}\hlstd{(}
  \hlcom{# the original adjacency matrix of the network to be partially observed}
  \hlstd{adjacencyMatrix,}
  \hlcom{# a character for the sampling design which defines the observation process}
  \hlstd{sampling,}
  \hlcom{# a set of parameters associated to the sampling design}
  \hlstd{parameters,}
  \hlcom{# an optional clustering membership vector required for block samplings}
  \hlkwc{clusters} \hlstd{=} \hlkwa{NULL}\hlstd{,}
  \hlcom{# an optional list of covariates that may influence the observation process}
  \hlkwc{covariates} \hlstd{=} \hlkwd{list}\hlstd{(),}
  \hlcom{# an optional function to compute similarities between node covariates}
  \hlkwc{similarity} \hlstd{= l1_similarity,}
  \hlcom{# an optional intercept term added in case of the presence of covariates}
  \hlkwc{intercept} \hlstd{=} \hlnum{0}\hlstd{)}
\end{alltt}
\end{kframe}
\end{knitrout}

Note that the dimension (or \texttt{length}) of \texttt{parameters} depends on the sampling design selected, as described in Section \ref{samp:def}. Argument \texttt{clusters} only needs to be specified for \texttt{"block-dyad"} and \texttt{"block-node"} sampling designs. Arguments \texttt{covariates} is by-default \texttt{list()}, \texttt{covarSimilarity} is set to the \texttt{l1\_similarity} function defined by $(x,y) \in \Rbb^d \times \Rbb^d \mapsto -|x-y|_{\ell^1} \in \Rbb^d$, and finally \texttt{intercept} is set to $0$. These last three arguments only need to be specified in the case of an SBM with covariate(s). Note that the intercept is not included in \texttt{parameters} and must be specified independently. {The output of \texttt{observeNetwork} is a \texttt{matrix} encoding the adjacency-matrix, with \texttt{NA} values for dyads not observed by the observation process, which can be provided as input to \texttt{estimateMissSBM}.}

\section{Illustration: the 2007 French political blogosphere} \label{sec:example}

This section illustrates the features of \pkg{missSBM} by conducting the analysis of a real-world network data. It is a sub-network of the French political blogosphere, extracted from a snapshot of over $1100$ blogs collected during a period preceding the 2007 French presidential election, and manually classified by the "Observatoire Pr\'esidentielle project" \citep[see][]{zanghi2008}. The network is composed of $194$ blogs representing nodes in the network and $1432$ edges indicating that at least one of the two blogs references the other. On top of \pkg{missSBM}, our analysis relies on \pkg{aricode} \citep{aricode} for computing clustering comparison measures and \pkg{tidyverse} for performing data manipulations and producing graphical outputs. The \pkg{igraph} package, imported by \pkg{missSBM}, is needed for basic graph manipulations. \new{The \pkg{future} package is used for parallel computing}. We also fix the seed for reproducibility:

\begin{knitrout}
\definecolor{shadecolor}{rgb}{0.969, 0.969, 0.969}\color{fgcolor}\begin{kframe}
\begin{alltt}
\hlkwd{library}\hlstd{(missSBM)}
\hlkwd{library}\hlstd{(aricode)}
\hlkwd{library}\hlstd{(tidyverse);} \hlkwd{theme_set}\hlstd{(}\hlkwd{theme_bw}\hlstd{())}
\hlkwd{library}\hlstd{(igraph)}
\hlkwd{library}\hlstd{(future)}
\hlkwd{set.seed}\hlstd{(}\hlnum{03052008}\hlstd{)}
\end{alltt}
\end{kframe}
\end{knitrout}

\new{We set our \texttt{future::plan} to \texttt{multicore} with 10 workers \footnote{Use \texttt{multisession} if you work on Windows or from Rstudio}.}

\begin{knitrout}
\definecolor{shadecolor}{rgb}{0.969, 0.969, 0.969}\color{fgcolor}\begin{kframe}
\begin{alltt}
\hlstd{future}\hlopt{::}\hlkwd{plan}\hlstd{(}\hlstr{"multicore"}\hlstd{,} \hlkwc{workers} \hlstd{=} \hlnum{10}\hlstd{)}
\end{alltt}
\end{kframe}
\end{knitrout}

The \texttt{frenchblog2007} data set is provided with \pkg{missSBM}\footnote{Earlier versions of this data set were available in packages \pkg{mixer} and \pkg{sand}.} as an igraph object. We extract the adjacency matrix corresponding to the blog network after removing the isolated nodes. We also extract the political party of each blog from the vertex attribute \proglang{party}, which gives us a natural classification of the nodes that could be used as a reference in our analyses.

\begin{knitrout}
\definecolor{shadecolor}{rgb}{0.969, 0.969, 0.969}\color{fgcolor}\begin{kframe}
\begin{alltt}
\hlkwd{data}\hlstd{(}\hlstr{"frenchblog2007"}\hlstd{,} \hlkwc{package} \hlstd{=} \hlstr{"missSBM"}\hlstd{)}
\hlstd{frenchblog2007} \hlkwb{<-}
  \hlkwd{delete_vertices}\hlstd{(frenchblog2007,} \hlkwd{which}\hlstd{(}\hlkwd{degree}\hlstd{(frenchblog2007)} \hlopt{==} \hlnum{0}\hlstd{))}
\hlstd{blog}  \hlkwb{<-} \hlkwd{as_adj}\hlstd{(frenchblog2007)}
\hlstd{party} \hlkwb{<-} \hlkwd{vertex.attributes}\hlstd{(frenchblog2007)}\hlopt{$}\hlstd{party}
\end{alltt}
\end{kframe}
\end{knitrout}

For this network, we explore models with a number of blocks varying from 1 to 18:
\begin{knitrout}
\definecolor{shadecolor}{rgb}{0.969, 0.969, 0.969}\color{fgcolor}\begin{kframe}
\begin{alltt}
\hlstd{blocks} \hlkwb{<-} \hlnum{1}\hlopt{:}\hlnum{18}
\end{alltt}
\end{kframe}
\end{knitrout}

\paragraph*{Standard SBM Estimation.} At this stage, the data set has no missing entry: every dyad and every node is observed. The adjacency matrix $\bY$ of the fully-observed network is stored in the variable \texttt{blog}. We first perform a standard SBM estimation on the fully observed network, including smoothing of the ICL. 

\begin{knitrout}
\definecolor{shadecolor}{rgb}{0.969, 0.969, 0.969}\color{fgcolor}\begin{kframe}
\begin{alltt}
\hlstd{sbm_full} \hlkwb{<-} \hlkwd{estimateMissSBM}\hlstd{(blog, blocks,} \hlstr{"node"}\hlstd{)}
\end{alltt}
\end{kframe}
\end{knitrout}

\new{We inspect the result of the optimization process by plotting the ELBO (variational lower bound of the log-likelihood) against the number of iterations in the V-EM algorithm, accumulated along all the numbers of blocks considered (Figure~\ref{fig:monitoring}). The ELBO is typically expected to increase with the number of blocks at the end of the successive optimizations, which is indeed the case here:}
\begin{knitrout}
\definecolor{shadecolor}{rgb}{0.969, 0.969, 0.969}\color{fgcolor}\begin{kframe}
\begin{alltt}
\hlkwd{plot}\hlstd{(sbm_full,} \hlkwc{type} \hlstd{=} \hlstr{"monitoring"}\hlstd{)}
\end{alltt}
\end{kframe}\begin{figure}[htbp!]

{\centering \includegraphics[width=.6\textwidth]{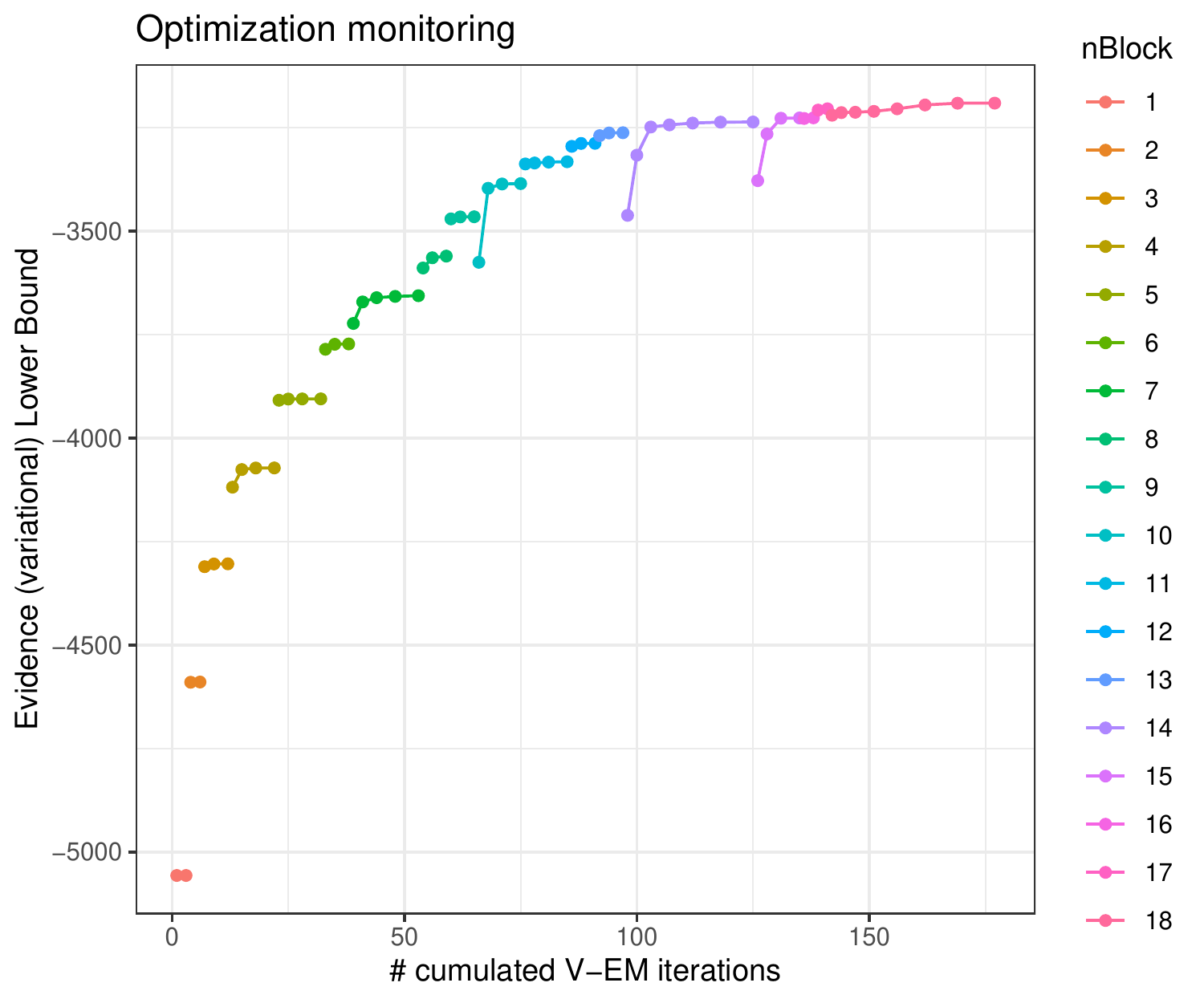} 

}

\caption{\new{\label{fig:monitoring}Evolution of the ELBO during the optimization for the successive numbers of blocks}}\label{fig:plot-monitoring-full}
\end{figure}

\end{knitrout}

\new{The ICL criterion is minimal for an SBM with 10 blocks:}
\begin{knitrout}
\definecolor{shadecolor}{rgb}{0.969, 0.969, 0.969}\color{fgcolor}\begin{kframe}
\begin{alltt}
\hlkwd{which.min}\hlstd{(sbm_full}\hlopt{$}\hlstd{ICL)}
\end{alltt}
\begin{verbatim}
## [1] 10
\end{verbatim}
\end{kframe}
\end{knitrout}

\new{The corresponding model is stored in the field \texttt{\$bestModel} of \texttt{sbm\_full} as an object with class \texttt{missSBM\_fit}. Printing this object results in a summary of the most important accessible fields and methods:}
\begin{knitrout}
\definecolor{shadecolor}{rgb}{0.969, 0.969, 0.969}\color{fgcolor}\begin{kframe}
\begin{alltt}
\hlstd{sbm_full}\hlopt{$}\hlstd{bestModel}
\end{alltt}
\begin{verbatim}
## missSBM-fit
## ==================================================================
## Structure for storing an SBM fitted under missing data condition  
## ==================================================================
## * Useful fields (first 2 special objects themselves with methods) 
##   $fittedSBM (the adjusted stochastic block model)                
##   $fittedSampling (the estimated sampling process)                
##   $imputedNetwork (the adjacency matrix with imputed values)      
##   $monitoring, $ICL, $loglik, $vExpec, $penalty                  
## * S3 methods                                                      
##   plot, coef, fitted, predict, print
\end{verbatim}
\end{kframe}
\end{knitrout}

In particular, one can access various parameters (e.g. block proportion/mixture parameters),
\begin{knitrout}
\definecolor{shadecolor}{rgb}{0.969, 0.969, 0.969}\color{fgcolor}\begin{kframe}
\begin{alltt}
\hlkwd{coef}\hlstd{(sbm_full}\hlopt{$}\hlstd{bestModel,} \hlkwc{type} \hlstd{=} \hlstr{"mixture"}\hlstd{)}
\end{alltt}
\begin{verbatim}
##  [1] 0.02741279 0.03589227 0.08965712 0.06583498 0.14612002 0.12372170
##  [7] 0.05670103 0.12990416 0.13922946 0.18552647
\end{verbatim}
\end{kframe}
\end{knitrout}

or plot diverse outputs, for instance a matrix view of the original network with columns and rows reordered according to the block memberships of the nodes, see Figure~\ref{fig:imputed}).
\begin{knitrout}
\definecolor{shadecolor}{rgb}{0.969, 0.969, 0.969}\color{fgcolor}\begin{kframe}
\begin{alltt}
\hlkwd{plot}\hlstd{(sbm_full}\hlopt{$}\hlstd{bestModel,} \hlkwc{dimLabels} \hlstd{=} \hlkwd{list}\hlstd{(}\hlkwc{row} \hlstd{=} \hlstr{"blogs"}\hlstd{,} \hlkwc{col} \hlstd{=} \hlstr{"blogs"}\hlstd{))}
\end{alltt}
\end{kframe}\begin{figure}[htbp!]

{\centering \includegraphics[width=.6\textwidth]{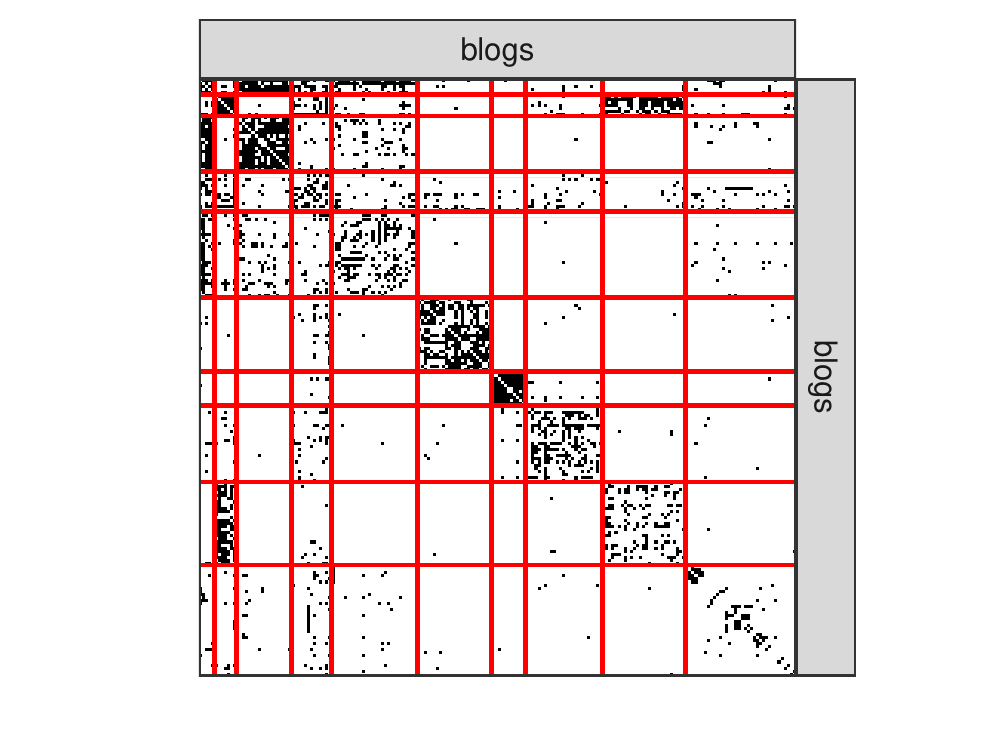} 

}

\caption{\new{\label{fig:imputed}Network data reorganized by the estimated block memberships}}\label{fig:plot-monitoring-best}
\end{figure}

\end{knitrout}

\paragraph*{{Partial Observation of an Existing Network.}} For illustrative purposes, we sample a sub-graph from the blog network to mimic missing data and create a new adjacency matrix with missing entries. Since data consists in interactions between blogs (who references who), it is natural to sample the network with a node-centered sampling design: the following code generates a realization of a $194 \times 194$ sampling matrix according to the \textit{block-node sampling}, where the blocks correspond to the clustering estimated by the SBM fitted on the full data set. The sampling rate is either low (0.2) in small blocks or high (0.8) in large blocks.

\begin{knitrout}
\definecolor{shadecolor}{rgb}{0.969, 0.969, 0.969}\color{fgcolor}\begin{kframe}
\begin{alltt}
\hlstd{samplingParameters} \hlkwb{<-}
  \hlkwd{ifelse}\hlstd{(sbm_full}\hlopt{$}\hlstd{bestModel}\hlopt{$}\hlstd{fittedSBM}\hlopt{$}\hlstd{blockProp} \hlopt{<} \hlnum{.1}\hlstd{,} \hlnum{0.2}\hlstd{,} \hlnum{.8}\hlstd{)}
\hlstd{blog_obs} \hlkwb{<-}
  \hlkwd{observeNetwork}\hlstd{(}
    \hlkwc{adjacencyMatrix} \hlstd{= blog,}
    \hlkwc{sampling}        \hlstd{=} \hlstr{"block-node"}\hlstd{,}
    \hlkwc{parameters}      \hlstd{= samplingParameters,}
    \hlkwc{clusters}        \hlstd{= sbm_full}\hlopt{$}\hlstd{bestModel}\hlopt{$}\hlstd{fittedSBM}\hlopt{$}\hlstd{memberships}
  \hlstd{)}
\end{alltt}
\end{kframe}
\end{knitrout}

\paragraph*{Estimation of a Partially Observed Network.} We now perform SBM inference under missing data conditions by fitting two types of model: first, the SBM under the MNAR \textit{block-node} sampling design, i.e., under the design that truly generated the missing entries; second, the SBM under the MCAR \textit{node} sampling design which basically performs inference only on the observed part of the network, neglecting the process that \new{generates} the missing values. The estimation is run on both models with the same settings as for the fully observed data. The ICL curve is smoothed with five iterations of forward-backward \new{exploration since the presence of missing values typically increases the chances for falling into local minima.}

\begin{knitrout}
\definecolor{shadecolor}{rgb}{0.969, 0.969, 0.969}\color{fgcolor}\begin{kframe}
\begin{alltt}
\hlstd{sbm_block} \hlkwb{<-}
  \hlkwd{estimateMissSBM}\hlstd{(blog_obs, blocks,} \hlstr{"block-node"}\hlstd{,} \hlkwc{control} \hlstd{=} \hlkwd{list}\hlstd{(}\hlkwc{iterates}\hlstd{=}\hlnum{5}\hlstd{))}
\hlstd{sbm_node} \hlkwb{<-}
  \hlkwd{estimateMissSBM}\hlstd{(blog_obs, blocks,} \hlstr{"node"}\hlstd{,} \hlkwc{control} \hlstd{=} \hlkwd{list}\hlstd{(}\hlkwc{iterates}\hlstd{=}\hlnum{5}\hlstd{))}
\end{alltt}
\end{kframe}
\end{knitrout}

\paragraph*{Sampling Design Comparison.} \new{We plot on Figure~\ref{fig:ICL_samplings} the ICL of the three models (\texttt{"fully observed"}, \texttt{"block-node"} sampling and \texttt{"node"} sampling) to show how it can be compared to select which sampling design fits at best the data:}

\begin{knitrout}
\definecolor{shadecolor}{rgb}{0.969, 0.969, 0.969}\color{fgcolor}\begin{kframe}
\begin{alltt}
\hlkwd{rbind}\hlstd{(}
  \hlkwd{tibble}\hlstd{(}\hlkwc{Q} \hlstd{= blocks,} \hlkwc{ICL} \hlstd{= sbm_node}\hlopt{$}\hlstd{ICL ,} \hlkwc{sampling} \hlstd{=}\hlstr{"node"}\hlstd{),}
  \hlkwd{tibble}\hlstd{(}\hlkwc{Q} \hlstd{= blocks,} \hlkwc{ICL} \hlstd{= sbm_block}\hlopt{$}\hlstd{ICL,} \hlkwc{sampling} \hlstd{=} \hlstr{"block-node"}\hlstd{),}
  \hlkwd{tibble}\hlstd{(}\hlkwc{Q} \hlstd{= blocks,} \hlkwc{ICL} \hlstd{= sbm_full}\hlopt{$}\hlstd{ICL ,} \hlkwc{sampling} \hlstd{=} \hlstr{"fully observed"}\hlstd{)}
\hlstd{)} \hlopt{%>%} \hlkwd{ggplot}\hlstd{(}\hlkwd{aes}\hlstd{(}\hlkwc{x} \hlstd{= Q,} \hlkwc{y} \hlstd{= ICL,} \hlkwc{color} \hlstd{= sampling))} \hlopt{+}
  \hlkwd{geom_line}\hlstd{()} \hlopt{+} \hlkwd{geom_point}\hlstd{()} \hlopt{+} \hlkwd{ggtitle}\hlstd{(}\hlstr{"Model Selection"}\hlstd{)} \hlopt{+}
  \hlkwd{labs}\hlstd{(}\hlkwc{x} \hlstd{=} \hlstr{"# blocks"}\hlstd{,} \hlkwc{y} \hlstd{=} \hlstr{"Integrated Classification Likelihood"}\hlstd{)}
\end{alltt}
\end{kframe}\begin{figure}

{\centering \includegraphics[width=.6\textwidth]{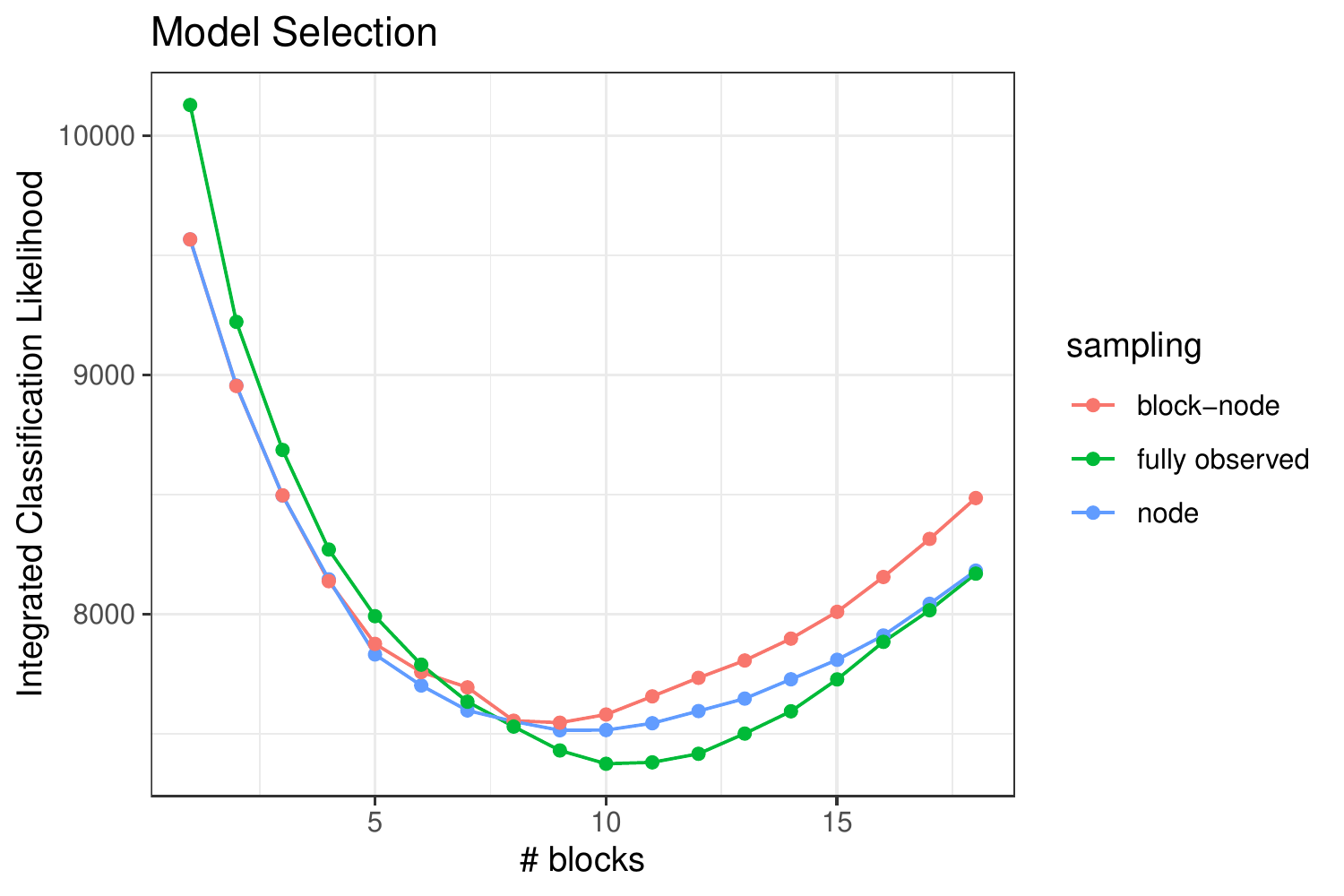} 

}

\caption{\new{\label{fig:ICL_samplings}ICL for models with fully observed data, block-node sampling and node sampling}}\label{fig:plot-ICLs-blog}
\end{figure}

\end{knitrout}

Note that the curves associated with the \textit{block-node sampling} and the \textit{node sampling} are quite close for small numbers of blocks (less than $10$) and then depart from each other: the choice of the sampling design is a tough question for the network data at play. We also represent the ICL curve for the collection of SBM estimated on the fully observed network: although the values of ICL cannot be compared between this model and the two obtained on the partially observed network (data sets are not the same), we underline that the numbers of blocks selected in the different cases remain comparable. A model with $10$ blocks is selected with the fully observed network, while a model with only $9$ blocks is chosen for the block-node sampling SBM. Indeed, due to the partial sampling, some blocks are less well represented than others, and it seems more likely to gather some blocks together considering the information available. Regarding the clustering obtained by the three variants, we compare them with the Adjusted Rand Index \citep[ARI,][]{rand1971} computed with the \pkg{aricode} package \citep{aricode}. We use the classification of the fully-observed SBM as a reference, since its clustering was used to sample the network with the block-node sampling design. We typically expect that a model relying on a better modeling of the missing values shall lead to a clustering closer to the reference. This is indeed the case when looking at the next piece of code, where it is shown that the ARI with the reference clustering is  higher for the block-sampling SBM  than for the MCAR node sampling SBM:

\begin{knitrout}
\definecolor{shadecolor}{rgb}{0.969, 0.969, 0.969}\color{fgcolor}\begin{kframe}
\begin{alltt}
\hlkwd{ARI}\hlstd{(sbm_block}\hlopt{$}\hlstd{bestModel}\hlopt{$}\hlstd{fittedSBM}\hlopt{$}\hlstd{memberships,}
    \hlstd{sbm_full}\hlopt{$}\hlstd{bestModel}\hlopt{$}\hlstd{fittedSBM}\hlopt{$}\hlstd{memberships)}
\end{alltt}
\begin{verbatim}
[1] 0.6570555
\end{verbatim}
\begin{alltt}
\hlkwd{ARI}\hlstd{(sbm_node}\hlopt{$}\hlstd{bestModel}\hlopt{$}\hlstd{fittedSBM}\hlopt{$}\hlstd{memberships ,}
    \hlstd{sbm_full}\hlopt{$}\hlstd{bestModel}\hlopt{$}\hlstd{fittedSBM}\hlopt{$}\hlstd{memberships)}
\end{alltt}
\begin{verbatim}
[1] 0.5436008
\end{verbatim}
\end{kframe}
\end{knitrout}

\paragraph{Extraction of the SBM with Block-Sampling Design.} The model that we finally retain is thus a block-sampling with 9 blocks.

\begin{knitrout}
\definecolor{shadecolor}{rgb}{0.969, 0.969, 0.969}\color{fgcolor}\begin{kframe}
\begin{alltt}
\hlstd{myModel} \hlkwb{<-} \hlstd{sbm_block}\hlopt{$}\hlstd{bestModel}
\end{alltt}
\end{kframe}
\end{knitrout}

\texttt{myModel} is an object with class \texttt{missSBM\_fit} with two special elements used for storing the results of the estimation of both the SBM (field \texttt{fittedSBM}) and the sampling design (\texttt{fittedSampling}). These two elements are special objects themselves with dedicated fields and methods which are recalled to the user thanks to the \texttt{print}/\texttt{show} methods:

\begin{knitrout}
\definecolor{shadecolor}{rgb}{0.969, 0.969, 0.969}\color{fgcolor}\begin{kframe}
\begin{alltt}
\hlstd{myModel}\hlopt{$}\hlstd{fittedSBM}
\end{alltt}
\begin{verbatim}
Simple Stochastic Block Model -- bernoulli variant
=====================================================================
Dimension = ( 194 ) - ( 9 ) blocks and no covariate(s).
=====================================================================
* Useful fields 
  $nbNodes, $modelName, $dimLabels, $nbBlocks, $nbCovariates, $nbDyads
  $blockProp, $connectParam, $covarParam, $covarList, $covarEffect 
  $expectation, $indMemberships, $memberships 
* R6 and S3 methods 
  $rNetwork, $rMemberships, $rEdges, plot, print, coef 
\end{verbatim}
\end{kframe}
\end{knitrout}

\begin{knitrout}
\definecolor{shadecolor}{rgb}{0.969, 0.969, 0.969}\color{fgcolor}\begin{kframe}
\begin{alltt}
\hlstd{myModel}\hlopt{$}\hlstd{fittedSampling}
\end{alltt}
\begin{verbatim}
block-node-model for network sampling
==================================================================
Structure for handling network sampling in missSBM.
==================================================================
* Useful fields 
  $type, $parameters, $df
  $penalty, $vExpec
\end{verbatim}
\end{kframe}
\end{knitrout}

\paragraph*{Representation and Validation.}

With \texttt{myModel}, we now have at hand a tool for analyzing the clustering of the French political blogosphere. The first output is the connectivity matrix of the network, which puts into light the community structure of the blogosphere. Indeed, it is revealed by a diagonal filled with high probabilities and off-diagonal with low probabilities. Thus, nodes (blogs) in blocks connect with a high probability with other nodes in the same block and with a low probability with nodes in other blocks. Such a network concentrates most of its edges between nodes of the same blocks. This can be seen by displaying the  probability of connection predicted by the SBM at the whole network scale (see Figure~\ref{fig:connectivity}):
\begin{knitrout}
\definecolor{shadecolor}{rgb}{0.969, 0.969, 0.969}\color{fgcolor}\begin{kframe}
\begin{alltt}
\hlkwd{plot}\hlstd{(myModel,} \hlkwc{type} \hlstd{=} \hlstr{"expected"}\hlstd{,} \hlkwc{dimLabels} \hlstd{=} \hlkwd{list}\hlstd{(}\hlkwc{row}\hlstd{=}\hlstr{"blogs"}\hlstd{,} \hlkwc{col}\hlstd{=}\hlstr{"blogs"}\hlstd{))}
\end{alltt}
\end{kframe}\begin{figure}[htbp!]

{\centering \includegraphics[width=.6\textwidth]{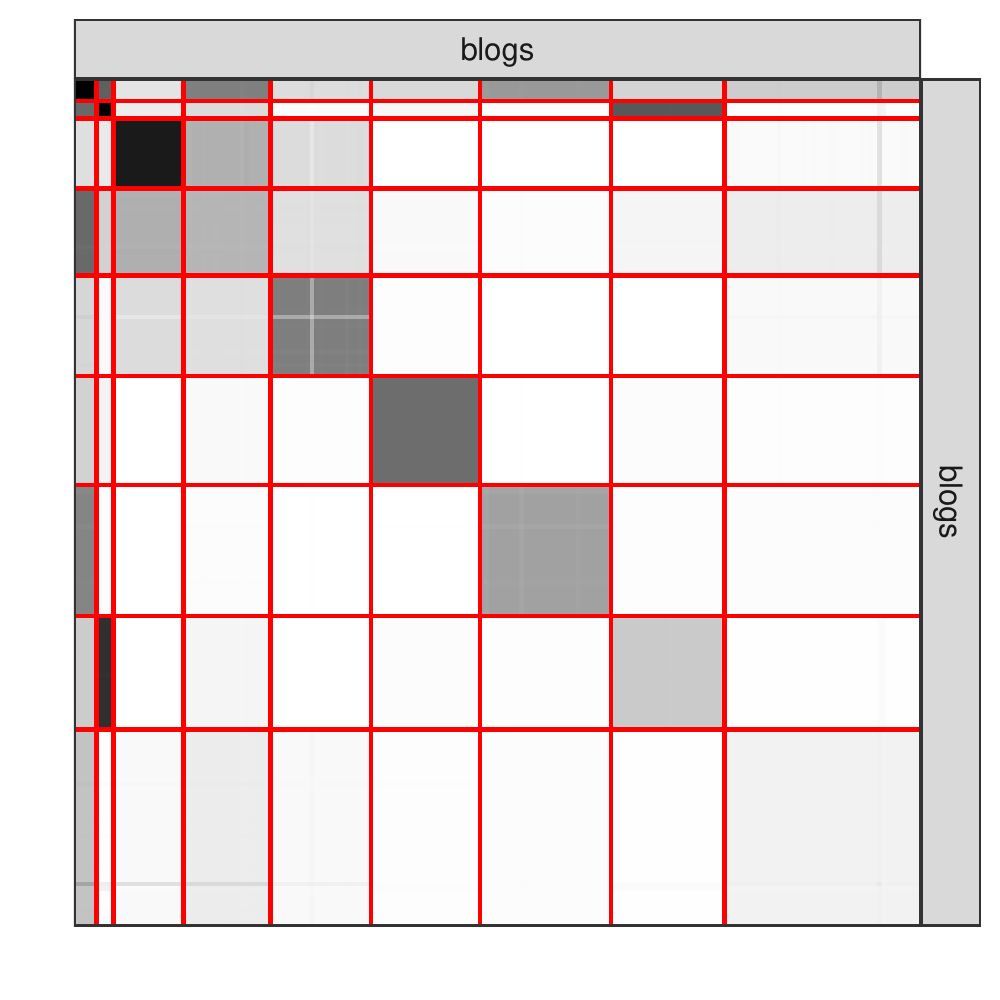} 

}

\caption{\label{fig:connectivity}Probabilities of connection predicted by the SBM with block-node sampling}\label{fig:connectivity-parameters}
\end{figure}

\end{knitrout}

For validation, we suggest comparing the clustering of the model with the node attribute corresponding to the political parties to which the blogs belong. First, we remark that the SBM fitted on missing entries carries a little bit less information regarding the political party than the SBM adjusted on the fully observed network:
\begin{knitrout}
\definecolor{shadecolor}{rgb}{0.969, 0.969, 0.969}\color{fgcolor}\begin{kframe}
\begin{alltt}
\hlkwd{ARI}\hlstd{(party, myModel}\hlopt{$}\hlstd{fittedSBM}\hlopt{$}\hlstd{memberships)}
\end{alltt}
\begin{verbatim}
[1] 0.4126328
\end{verbatim}
\begin{alltt}
\hlkwd{ARI}\hlstd{(party, sbm_full}\hlopt{$}\hlstd{bestModel}\hlopt{$}\hlstd{fittedSBM}\hlopt{$}\hlstd{memberships)}
\end{alltt}
\begin{verbatim}
[1] 0.463709
\end{verbatim}
\end{kframe}
\end{knitrout}

A more detailed comparison between blocks inferred by the SBM and political parties is reported in Figure \ref{fig:alluvialPlot} with an alluvial diagram.

\begin{knitrout}
\definecolor{shadecolor}{rgb}{0.969, 0.969, 0.969}\color{fgcolor}\begin{figure}[htbp!]

{\centering \includegraphics[width=.8\textwidth]{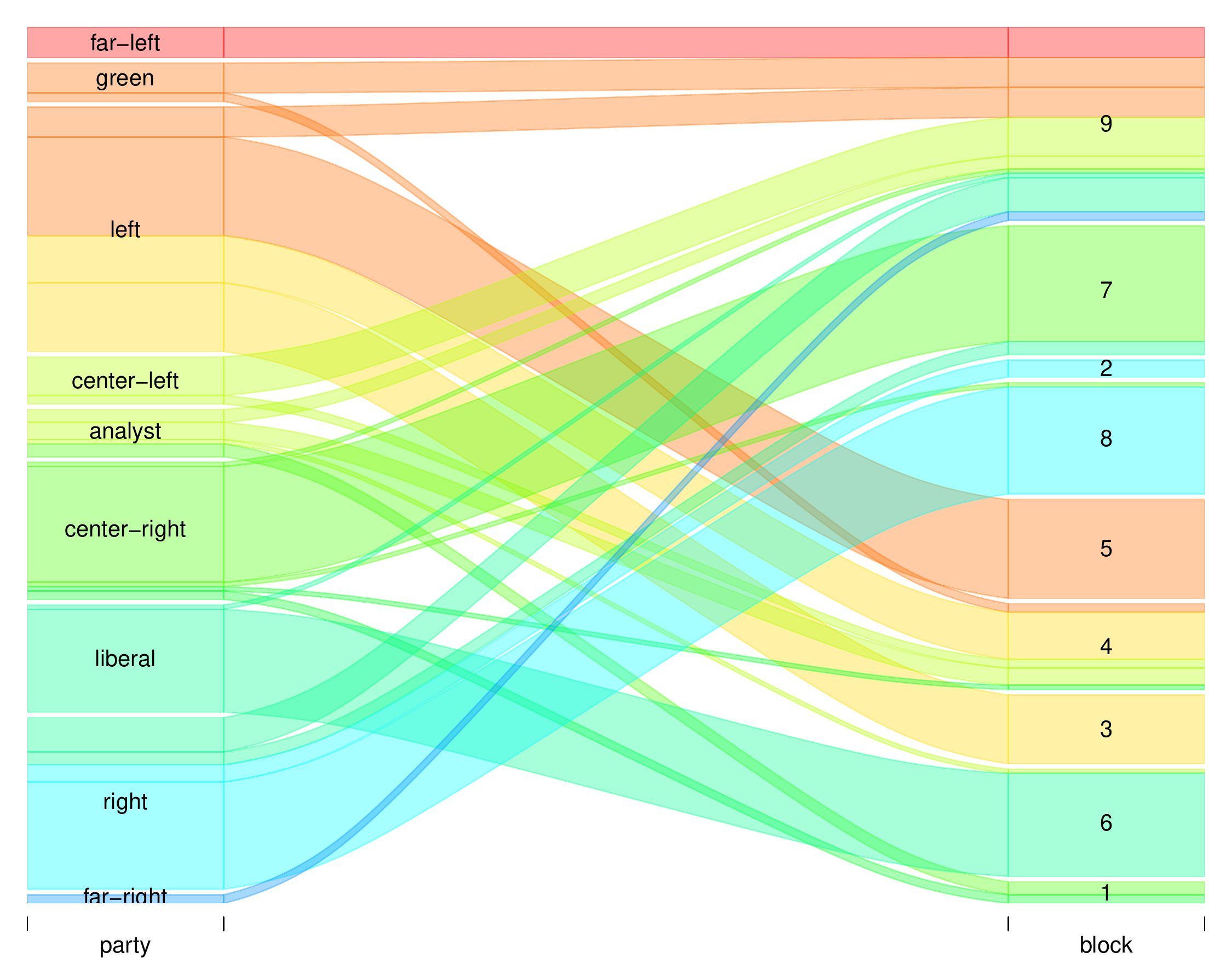} 

}

\caption{\label{fig:alluvialPlot}Alluvial plot between block-node sampling clustering and political parties.}\label{fig:Alluvial-plot}
\end{figure}

\end{knitrout}

Also remember that \pkg{missSBM} performs imputation of the missing dyads in  the adjacency matrix. Thus, we can compare the imputed values with the values of the dyad in the fully observed network to validate the performance of our approach. Using the \proglang{R} package \pkg{pROC} \citep{pROC}, we check the quality of the imputation. We replicate this experiment 500 times with a sampling rate varying between $\approx 0.4$ and $\approx 0.9$ (always with block-sampling design), fixing the number of blocks to the best one found on the fully observed network. The Area Under the Curve (AUC) is plotted in Figure \ref{fig:AUC} against the sampling rate, showing the robustness and the good performance of the imputation method.

\begin{knitrout}
\definecolor{shadecolor}{rgb}{0.969, 0.969, 0.969}\color{fgcolor}\begin{kframe}
\begin{alltt}
\hlkwd{library}\hlstd{(pROC)}
\hlkwd{library}\hlstd{(parallel)}
\hlstd{cl0}     \hlkwb{<-} \hlstd{sbm_full}\hlopt{$}\hlstd{bestModel}\hlopt{$}\hlstd{fittedSBM}\hlopt{$}\hlstd{memberships}
\hlstd{nBlocks} \hlkwb{<-} \hlstd{sbm_full}\hlopt{$}\hlstd{bestModel}\hlopt{$}\hlstd{fittedSBM}\hlopt{$}\hlstd{nbBlocks}
\hlstd{future}\hlopt{::}\hlkwd{plan}\hlstd{(}\hlstr{"sequential"}\hlstd{)}
\hlstd{res_auc} \hlkwb{<-} \hlkwd{mclapply}\hlstd{(}\hlnum{1}\hlopt{:}\hlnum{500}\hlstd{,} \hlkwa{function}\hlstd{(}\hlkwc{i}\hlstd{) \{}
  \hlstd{subGraph}   \hlkwb{<-} \hlkwd{observeNetwork}\hlstd{(blog,} \hlstr{"block-node"}\hlstd{,} \hlkwd{runif}\hlstd{(nBlocks), cl0)}
  \hlstd{missing}    \hlkwb{<-} \hlkwd{which}\hlstd{(}\hlkwd{as.matrix}\hlstd{(}\hlkwd{is.na}\hlstd{(subGraph)))}
  \hlstd{true_dyads} \hlkwb{<-} \hlstd{blog[missing]}
  \hlstd{sbm_block}  \hlkwb{<-} \hlkwd{estimateMissSBM}\hlstd{(subGraph, nBlocks,} \hlstr{"block-node"}\hlstd{,}
                    \hlkwc{control} \hlstd{=} \hlkwd{list}\hlstd{(}\hlkwc{cores} \hlstd{=} \hlnum{1}\hlstd{,} \hlkwc{trace} \hlstd{=} \hlnum{0}\hlstd{))}
  \hlstd{imputed_dyads} \hlkwb{<-} \hlstd{sbm_block}\hlopt{$}\hlstd{bestModel}\hlopt{$}\hlstd{imputedNetwork[missing]}
  \hlkwd{c}\hlstd{(}\hlkwc{rate} \hlstd{=} \hlnum{1} \hlopt{-} \hlkwd{length}\hlstd{(missing)}\hlopt{/}\hlkwd{length}\hlstd{(blog),}
    \hlkwc{auc}  \hlstd{=} \hlkwd{auc}\hlstd{(true_dyads, imputed_dyads,} \hlkwc{quiet} \hlstd{=} \hlnum{TRUE}\hlstd{))}
\hlstd{\},} \hlkwc{mc.cores} \hlstd{=} \hlnum{10}\hlstd{)}

\hlstd{purrr}\hlopt{::}\hlkwd{reduce}\hlstd{(res_auc, rbind)} \hlopt{%>%} \hlkwd{as.data.frame}\hlstd{()} \hlopt{%>%}
  \hlkwd{ggplot}\hlstd{()} \hlopt{+} \hlkwd{aes}\hlstd{(}\hlkwc{x} \hlstd{= rate,} \hlkwc{y} \hlstd{= auc)} \hlopt{+} \hlkwd{geom_point}\hlstd{(}\hlkwc{size} \hlstd{=} \hlnum{0.25}\hlstd{)} \hlopt{+}
  \hlkwd{geom_smooth}\hlstd{(}\hlkwc{method} \hlstd{=} \hlstr{"lm"}\hlstd{,} \hlkwc{formula} \hlstd{= y} \hlopt{~} \hlstd{x,} \hlkwc{se} \hlstd{=} \hlnum{FALSE}\hlstd{)} \hlopt{+}
  \hlkwd{labs}\hlstd{(}\hlkwc{x} \hlstd{=} \hlstr{"sampling rate"}\hlstd{,} \hlkwc{y} \hlstd{=} \hlstr{"Area under the ROC curve"}\hlstd{)}
\end{alltt}
\end{kframe}
\end{knitrout}

\begin{knitrout}
\definecolor{shadecolor}{rgb}{0.969, 0.969, 0.969}\color{fgcolor}\begin{figure}[htbp!]

{\centering \includegraphics[width=.6\textwidth]{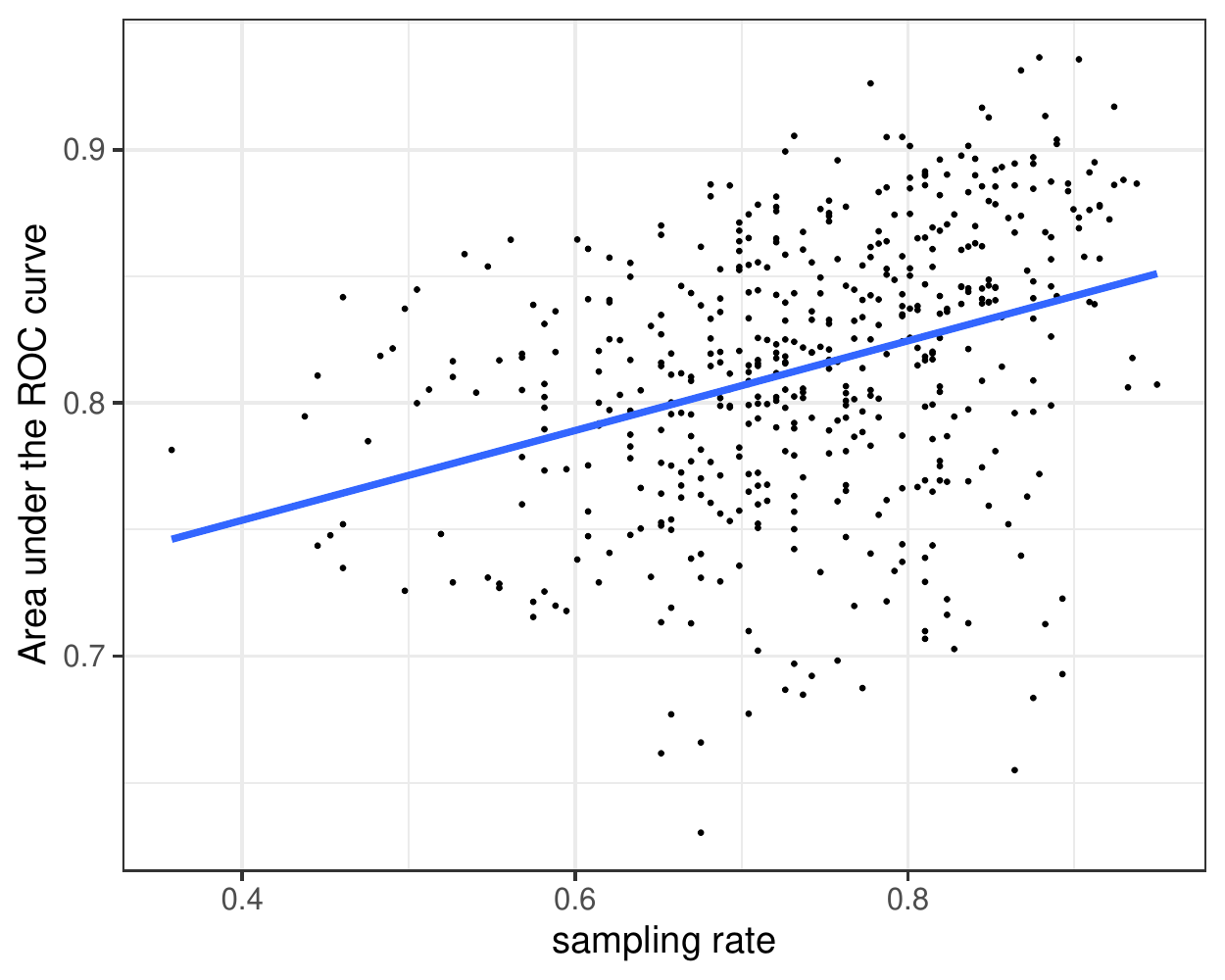} 

}

\caption{\label{fig:AUC}Area Under the Curve (AUC) of the imputation as a function of the sampling rate.}\label{fig:AUC-sampling-show}
\end{figure}

\end{knitrout}

\paragraph*{{Dealing with Covariates.}}

In order to illustrate how we can deal with covariates in the observation process, we now consider a sampling which depends on the political parties of the blog. For illustrative purposes, we extract a sub-graph that only contains the nodes whose political party is either \texttt{left} or \texttt{right}:
\begin{knitrout}
\definecolor{shadecolor}{rgb}{0.969, 0.969, 0.969}\color{fgcolor}\begin{kframe}
\begin{alltt}
\hlstd{blog_subgraph} \hlkwb{<-}
  \hlstd{frenchblog2007} \hlopt{%>%}
  \hlstd{igraph}\hlopt{::}\hlkwd{induced_subgraph}\hlstd{(}\hlkwd{V}\hlstd{(frenchblog2007)}\hlopt{$}\hlstd{party} \hlopt{%in%} \hlkwd{c}\hlstd{(} \hlstr{"right"}\hlstd{,} \hlstr{"left"}\hlstd{))}
\hlstd{blog_subgraph} \hlkwb{<-}
  \hlkwd{delete_vertices}\hlstd{(blog_subgraph,} \hlkwd{which}\hlstd{(}\hlkwd{degree}\hlstd{(blog_subgraph)} \hlopt{==} \hlnum{0}\hlstd{))}
\end{alltt}
\end{kframe}
\end{knitrout}

\new{The sub-graph is given in Figure~\ref{fig:subgraph}, generated with the following piece of code:}

\begin{knitrout}
\definecolor{shadecolor}{rgb}{0.969, 0.969, 0.969}\color{fgcolor}\begin{kframe}
\begin{alltt}
\hlkwd{plot}\hlstd{(blog_subgraph,} \hlkwc{vertex.shape}\hlstd{=}\hlstr{"none"}\hlstd{,} \hlkwc{vertex.label}\hlstd{=}\hlkwd{V}\hlstd{(blog_subgraph)}\hlopt{$}\hlstd{party,}
                    \hlkwc{vertex.label.color} \hlstd{=} \hlstr{"steel blue"}\hlstd{,} \hlkwc{vertex.label.font}\hlstd{=}\hlnum{1.5}\hlstd{,}
                    \hlkwc{vertex.label.cex}\hlstd{=}\hlnum{.6}\hlstd{,} \hlkwc{edge.color}\hlstd{=}\hlstr{"gray70"}\hlstd{,} \hlkwc{edge.width} \hlstd{=} \hlnum{1}\hlstd{)}
\end{alltt}
\end{kframe}\begin{figure}[htbp!]

{\centering \includegraphics[width=0.65\linewidth]{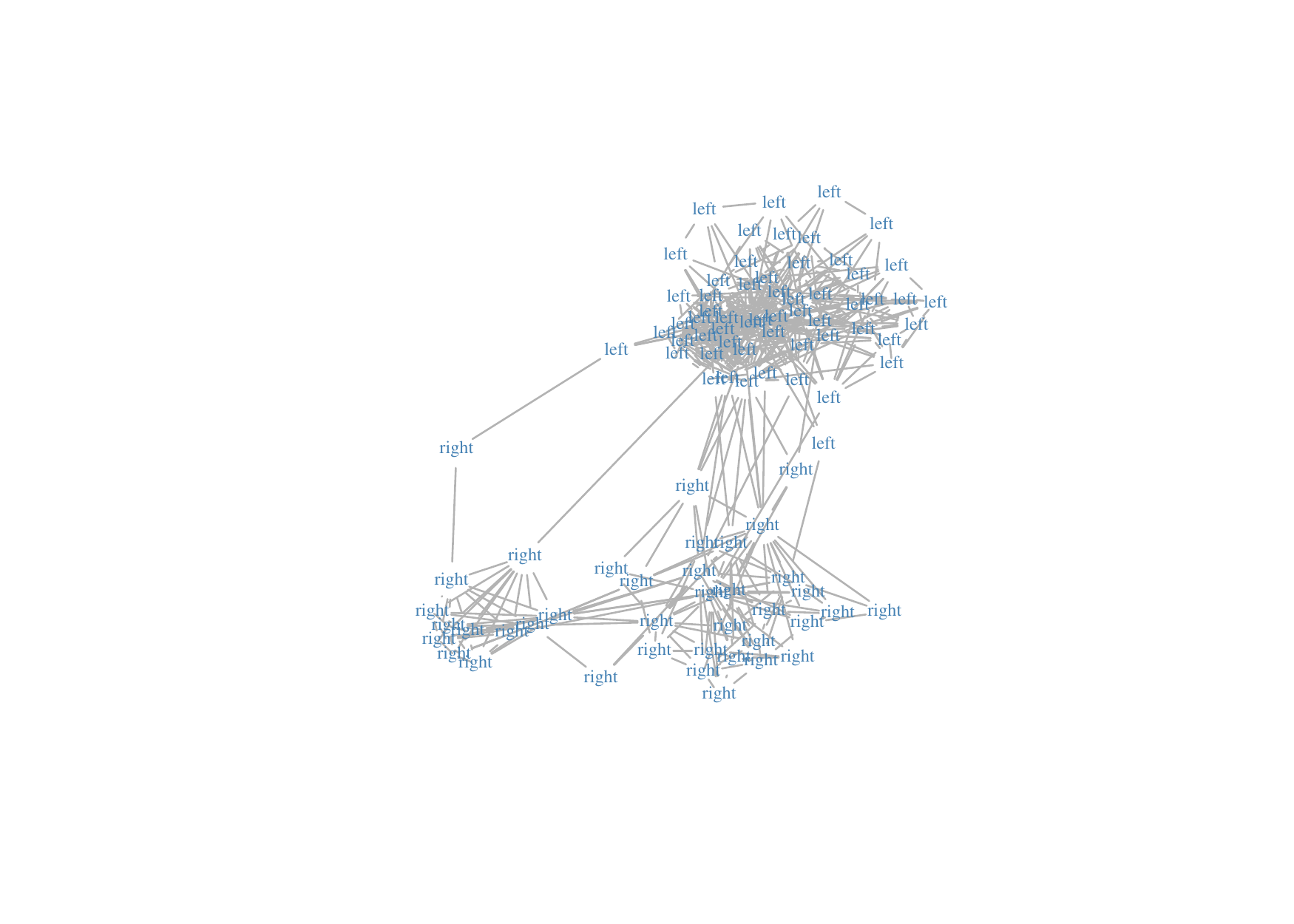} 

}

\caption{\new{\label{fig:subgraph}Subnetwork extracted from the French political blogosphere (only blogs with attribute party in $\{\mathrm{left,right}\}$ were kept)}}\label{fig:plot-subgraph}
\end{figure}

\end{knitrout}

We then build a simple binary covariate on nodes indicating its party (1/0 = left/right).

\begin{knitrout}
\definecolor{shadecolor}{rgb}{0.969, 0.969, 0.969}\color{fgcolor}\begin{kframe}
\begin{alltt}
\hlstd{dummy_party} \hlkwb{<-} \hlstd{(}\hlkwd{V}\hlstd{(blog_subgraph)}\hlopt{$}\hlstd{party} \hlopt{==} \hlstr{"left"}\hlstd{)} \hlopt{*} \hlnum{1}
\end{alltt}
\end{kframe}
\end{knitrout}

Now, the observation process of the network is assumed to depend on this covariate so that a node belonging to the \texttt{left} party is more likely to be observed than a node belonging to the \texttt{right} party:

\begin{knitrout}
\definecolor{shadecolor}{rgb}{0.969, 0.969, 0.969}\color{fgcolor}\begin{kframe}
\begin{alltt}
\hlstd{blog_subgraph_obs} \hlkwb{<-}
  \hlstd{blog_subgraph} \hlopt{%>%} \hlkwd{as_adj}\hlstd{()} \hlopt{%>%}
  \hlkwd{observeNetwork}\hlstd{(}\hlkwc{sampling}\hlstd{=}\hlstr{"covar-node"}\hlstd{,} \hlkwc{parameters} \hlstd{=} \hlnum{10}\hlstd{,}
                 \hlkwc{covariates} \hlstd{=} \hlkwd{list}\hlstd{(dummy_party))}
\hlstd{blocks} \hlkwb{<-} \hlnum{2}\hlopt{:}\hlnum{8}
\hlstd{future}\hlopt{::}\hlkwd{plan}\hlstd{(}\hlstr{"multicore"}\hlstd{,} \hlkwc{workers} \hlstd{=} \hlnum{10}\hlstd{)}
\end{alltt}
\end{kframe}
\end{knitrout}

Since the covariate is nodal, a nodal observation process is considered. When it comes to take into account the effect of this covariate on the probability of connection between two nodes in the SBM as in Equation \eqref{eq:SBM_covariates}, the covariate has to be transferred at the dyad level. This consists in computing an $n\times n$ matrix whose elements are equal to one if the corresponding two nodes belong to the same political party, zero otherwise. This matrix computation is directly handled in \pkg{missSBM}, when the option \texttt{useCov} is \texttt{TRUE}. \newline

From the observed network \texttt{blog\_subgraph\_obs}, four modeling choices (labeled $i$ to $iv$) are possible whether the covariate is taken into account in the SBM or not, and whether the sampling is set as depending on the covariate or not. For the latter choice, we know that the sampling which depends on the covariate is more appropriate but this information is generally not available \textit{a priori}. We then run the estimation in these four scenarios below and print the estimated parameters related to the SBM and the sampling.

\begin{enumerate}[i)]
\item Take the covariate into account in both the sampling and the SBM:
\begin{knitrout}
\definecolor{shadecolor}{rgb}{0.969, 0.969, 0.969}\color{fgcolor}\begin{kframe}
\begin{alltt}
\hlstd{sbm_covar1} \hlkwb{<-} \hlkwd{estimateMissSBM}\hlstd{(blog_subgraph_obs, blocks,}
  \hlstr{"covar-node"}\hlstd{,} \hlkwc{covariates} \hlstd{=} \hlkwd{list}\hlstd{(dummy_party),}
  \hlkwc{control} \hlstd{=} \hlkwd{list}\hlstd{(}\hlkwc{useCov} \hlstd{=} \hlnum{TRUE}\hlstd{,} \hlkwc{iterates} \hlstd{=} \hlnum{2}\hlstd{))}
\hlstd{sbm_covar1}\hlopt{$}\hlstd{bestModel}\hlopt{$}\hlstd{fittedSampling}\hlopt{$}\hlstd{parameters} \hlcom{# sampling parameters}
\hlstd{sbm_covar1}\hlopt{$}\hlstd{bestModel}\hlopt{$}\hlstd{fittedSBM}\hlopt{$}\hlstd{covarParam}      \hlcom{# regression parameter}
\end{alltt}
\end{kframe}
\end{knitrout}

\begin{knitrout}
\definecolor{shadecolor}{rgb}{0.969, 0.969, 0.969}\color{fgcolor}\begin{kframe}
\begin{verbatim}
## [1] -0.3629055  2.9469030
## [1] 4.945801
\end{verbatim}
\end{kframe}
\end{knitrout}

\item Take the covariate into account in the sampling only: 
\begin{knitrout}
\definecolor{shadecolor}{rgb}{0.969, 0.969, 0.969}\color{fgcolor}\begin{kframe}
\begin{alltt}
\hlstd{sbm_covar2} \hlkwb{<-} \hlkwd{estimateMissSBM}\hlstd{(blog_subgraph_obs, blocks,}
  \hlstr{"covar-node"}\hlstd{,} \hlkwc{covariates} \hlstd{=} \hlkwd{list}\hlstd{(dummy_party),}
  \hlkwc{control} \hlstd{=} \hlkwd{list}\hlstd{(}\hlkwc{useCov} \hlstd{=} \hlnum{FALSE}\hlstd{,} \hlkwc{iterates} \hlstd{=} \hlnum{2}\hlstd{))}
\hlstd{sbm_covar2}\hlopt{$}\hlstd{bestModel}\hlopt{$}\hlstd{fittedSampling}\hlopt{$}\hlstd{parameters} \hlcom{# sampling parameters}
\end{alltt}
\end{kframe}
\end{knitrout}

\begin{knitrout}
\definecolor{shadecolor}{rgb}{0.969, 0.969, 0.969}\color{fgcolor}\begin{kframe}
\begin{verbatim}
## [1] -0.3629055  2.9469030
\end{verbatim}
\end{kframe}
\end{knitrout}

\item Take the covariate into account in the SBM only: 
\begin{knitrout}
\definecolor{shadecolor}{rgb}{0.969, 0.969, 0.969}\color{fgcolor}\begin{kframe}
\begin{alltt}
\hlstd{sbm_covar3} \hlkwb{<-} \hlkwd{estimateMissSBM}\hlstd{(blog_subgraph_obs, blocks,}
  \hlstr{"node"}\hlstd{,} \hlkwc{covariates} \hlstd{=} \hlkwd{list}\hlstd{(dummy_party),}
  \hlkwc{control} \hlstd{=} \hlkwd{list}\hlstd{(}\hlkwc{useCov} \hlstd{=} \hlnum{TRUE}\hlstd{,} \hlkwc{iterates} \hlstd{=} \hlnum{2}\hlstd{))}
\hlstd{sbm_covar3}\hlopt{$}\hlstd{bestModel}\hlopt{$}\hlstd{fittedSampling}\hlopt{$}\hlstd{parameters} \hlcom{# sampling parameter}
\hlstd{sbm_covar3}\hlopt{$}\hlstd{bestModel}\hlopt{$}\hlstd{fittedSBM}\hlopt{$}\hlstd{covarParam}      \hlcom{# regression parameter}
\end{alltt}
\end{kframe}
\end{knitrout}

\begin{knitrout}
\definecolor{shadecolor}{rgb}{0.969, 0.969, 0.969}\color{fgcolor}\begin{kframe}
\begin{verbatim}
## [1] 0.71875
## [1] 4.945801
\end{verbatim}
\end{kframe}
\end{knitrout}

\item Ignore the covariate in both the sampling and the SBM:
\begin{knitrout}
\definecolor{shadecolor}{rgb}{0.969, 0.969, 0.969}\color{fgcolor}\begin{kframe}
\begin{alltt}
\hlstd{sbm_covar4} \hlkwb{<-} \hlkwd{estimateMissSBM}\hlstd{(blog_subgraph_obs, blocks,}
  \hlstr{"node"}\hlstd{,} \hlkwc{control} \hlstd{=} \hlkwd{list}\hlstd{(}\hlkwc{useCov} \hlstd{=} \hlnum{FALSE}\hlstd{,} \hlkwc{iterates} \hlstd{=} \hlnum{2}\hlstd{))}
\hlstd{sbm_covar4}\hlopt{$}\hlstd{bestModel}\hlopt{$}\hlstd{fittedSampling}\hlopt{$}\hlstd{parameters} \hlcom{# sampling parameter}
\end{alltt}
\end{kframe}
\end{knitrout}

\begin{knitrout}
\definecolor{shadecolor}{rgb}{0.969, 0.969, 0.969}\color{fgcolor}\begin{kframe}
\begin{verbatim}
## [1] 0.71875
\end{verbatim}
\end{kframe}
\end{knitrout}

\end{enumerate}

For models $i)$ and $ii)$, we notice that the estimates of the sampling parameters are quite accurate. For models $iii)$ and $iv)$, the estimation of the unique sampling parameter boils down to the proportion of observed nodes. For models $i)$ and $iii)$, the estimate of the parameter $\beta$ in Equation~\eqref{eq:SBM_covariates} is positive and quite high. Thus, we conclude that the probability of connection between two nodes is strengthened by the fact that the nodes belong to the same party.

Figure~\ref{fig:ICLcovar} shows clearly that the \texttt{"covar-node"} sampling is uncovered from the data since the ICL criterion favors this sampling no matter if the covariate is taken into account in the SBM or not. The ICL criterion also points out that the models including the covariate effect in the SBM are better.
\begin{knitrout}
\definecolor{shadecolor}{rgb}{0.969, 0.969, 0.969}\color{fgcolor}\begin{kframe}
\begin{alltt}
\hlkwd{rbind}\hlstd{(}
 \hlkwd{tibble}\hlstd{(}\hlkwc{Q}\hlstd{=blocks,} \hlkwc{ICL}\hlstd{=sbm_covar1}\hlopt{$}\hlstd{ICL,} \hlkwc{sampling}\hlstd{=}\hlstr{"covar-node"}\hlstd{,} \hlkwc{useCov}\hlstd{=}\hlstr{'true'} \hlstd{),}
 \hlkwd{tibble}\hlstd{(}\hlkwc{Q}\hlstd{=blocks,} \hlkwc{ICL}\hlstd{=sbm_covar2}\hlopt{$}\hlstd{ICL,} \hlkwc{sampling}\hlstd{=}\hlstr{"covar-node"}\hlstd{,} \hlkwc{useCov}\hlstd{=}\hlstr{'false'}\hlstd{),}
 \hlkwd{tibble}\hlstd{(}\hlkwc{Q}\hlstd{=blocks,} \hlkwc{ICL}\hlstd{=sbm_covar3}\hlopt{$}\hlstd{ICL,} \hlkwc{sampling}\hlstd{=}\hlstr{"node"}      \hlstd{,} \hlkwc{useCov}\hlstd{=}\hlstr{'true'} \hlstd{),}
 \hlkwd{tibble}\hlstd{(}\hlkwc{Q}\hlstd{=blocks,} \hlkwc{ICL}\hlstd{=sbm_covar4}\hlopt{$}\hlstd{ICL,} \hlkwc{sampling}\hlstd{=}\hlstr{"node"}      \hlstd{,} \hlkwc{useCov}\hlstd{=}\hlstr{'false'}\hlstd{)}
\hlstd{)} \hlopt{%>%} \hlkwd{ggplot}\hlstd{(}\hlkwd{aes}\hlstd{(}\hlkwc{x} \hlstd{= Q,} \hlkwc{y} \hlstd{= ICL,} \hlkwc{color} \hlstd{= sampling,} \hlkwc{shape} \hlstd{= useCov))}  \hlopt{+}
  \hlkwd{geom_line}\hlstd{()} \hlopt{+} \hlkwd{geom_point}\hlstd{()} \hlopt{+} \hlkwd{labs}\hlstd{(}\hlkwc{x} \hlstd{=} \hlstr{"#blocks"}\hlstd{,} \hlkwc{y} \hlstd{=} \hlstr{"ICLs"}\hlstd{)}
\end{alltt}
\end{kframe}\begin{figure}[htbp!]

{\centering \includegraphics[width=.6\textwidth]{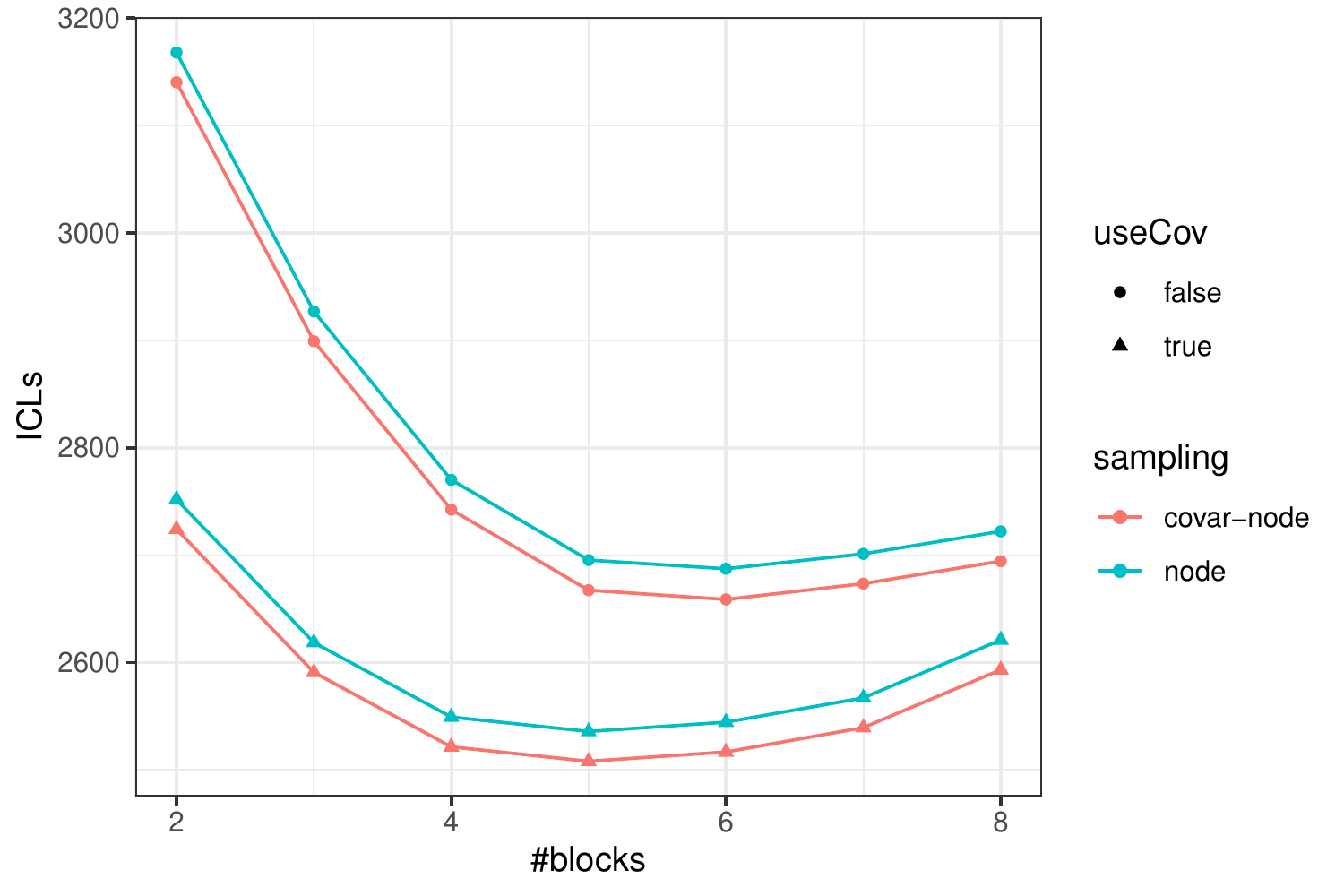} 

}

\caption{\label{fig:ICLcovar}ICL criterion for different numbers of blocks under the four models which make different use of the covariate political party.}\label{fig:plot-ICLs-covar}
\end{figure}

\end{knitrout}

Finally, we compare the clustering obtained with the four models together by computing pairwise ARIs. We also compare them with the clustering obtained by the SBM fitted on the fully observed network, adjusted as follows:
\begin{knitrout}
\definecolor{shadecolor}{rgb}{0.969, 0.969, 0.969}\color{fgcolor}\begin{kframe}
\begin{alltt}
\hlstd{sbm_covar_full} \hlkwb{<-} \hlkwd{as_adj}\hlstd{(blog_subgraph)} \hlopt{%>%}
  \hlkwd{estimateMissSBM}\hlstd{(blocks,} \hlstr{"node"}\hlstd{,} \hlkwc{covariates} \hlstd{=}  \hlkwd{list}\hlstd{(dummy_party))}
\end{alltt}
\end{kframe}
\end{knitrout}

For the sake of clarity, the ARIs are displayed in Table \ref{tab:covarARIs}. We obtain the same clustering with model $i)$ and $iii)$, which is the closest to the one obtained with the fully observed network. This is expected since these models also take into account the effect of the covariate.

\begin{knitrout}
\definecolor{shadecolor}{rgb}{0.969, 0.969, 0.969}\color{fgcolor}\begin{table}[!h]

\caption{\label{tab:ARI-covar}\label{tab:covarARIs}Clustering comparison with adjusted Rand indices (ARIs) between models taking the covariate into account (models $i)$, $ii)$, $iii)$, $iv)$ and model with fully observed network).}
\centering
\begin{tabular}[t]{lrrrr}
\toprule
  & i & ii & iii & iv\\
\midrule
Full & 0.64 & 0.48 & 0.64 & 0.44\\
i & NA & 0.42 & 1.00 & 0.39\\
ii & NA & NA & 0.42 & 0.95\\
iii & NA & NA & NA & 0.39\\
\bottomrule
\end{tabular}
\end{table}

\end{knitrout}


\section{Discussion}

The \proglang{R} package \pkg{missSBM} enables the estimation of an SBM from partially observed binary networks even for some observation processes which generate MNAR data.

Although version \texttt{1.0.0} of \pkg{missSBM} only deals with binary networks, we deploy a structure that lets the possibility to easily include other variants in the future. In particular, extending \pkg{missSBM} to weighted SBM where the distribution on the edges belongs to the exponential family \citep[e.g. Poisson distribution or Gaussian distribution, see][]{mariadassou2010} should be straightforward in the MAR case. To pave the way of such a generalization, \pkg{missSBM} relies on the package \pkg{sbm} which is designed to offer a collection of methods and algorithms to manipulate more general SBM, yet \emph{in the absence of missing data, or at least, with imputed data}. Hence, \pkg{missSBM} could take advantage in the future of improvements and new advances in \pkg{sbm}, while focusing on the handling of missing data specifically, by dealing only with the modeling of the observation process and the imputation of the missing entries. {The modular object-oriented coding of \pkg{missSBM} also allows the developer to make it easily evolving in the way it can take into account the missing data: new observation processes could be incorporated by providing new sampling design which should contain the functions corresponding to the sampling specific steps in the VEM algorithm. Other dependency structures between the SBM and the covariates could also be modeled and integrated. For example, the latent variables $\bZ$ could directly depend on the covariates $\bX$ instead of the adjacency matrix $\bY$.

Some recent work is concerned with boosting the speed of the variational inference \citep[see e.g.][]{blei2017variational}. This work could be adapted in the SBM context with or without missing data in order to enable our package to deal with larger networks.
\new{Resorting to minorization-maximization algorithms within the Variational E-step as in \citet{vu2013model} could also be interesting for speeding up the algorithm.}
Finally, a common drawback of the variational inference is to provide too narrow standard errors for the estimated parameters \citep{westling2015beyond}. Moreover, there is no uncertainty measure on the stability of the node clustering. We consider it as future work to provide the user with confidence intervals and stability measures of the clustering based on resampling techniques such as the bootstrap.

\section*{Acknowledgments}

The authors thank all members of MIRES group for fruitful discussions on network sampling designs.
This work is supported by public grants overseen by the French National research Agency (ANR) as part of the "Investissement d'Avenir" program, through the "IDI 2017" project funded by the IDEX Paris-Saclay, ANR-11-IDEX-0003-02, and second by the "EcoNet" project, ANR-18-CE02-0010.


\bibliography{sbmsampling}

\end{document}